\documentclass[letterpaper]{aa}
\usepackage{natbib}\usepackage{graphicx}      
\bibpunct{(}{)}{;}{a}{}{,}                

\newcommand{\kms}{km\,s$^{-1}$} \newcommand{\sqcm}{cm$^{-2}$}  
\newcommand{\hi}{\ion{H}{i}}    \newcommand{\hw}{\ion{H}{ii}} 
\newcommand{\os}{\ion{O}{vi}}   \newcommand{\cf}{\ion{C}{iv}} 
\newcommand{\sif}{\ion{Si}{iv}} \newcommand{\nf}{\ion{N}{v}}
\newcommand{\sit}{\ion{Si}{iii}}\newcommand{\ct}{\ion{C}{iii}}   
\newcommand{\siw}{\ion{Si}{ii}} \newcommand{\cw}{\ion{C}{ii}}   
\newcommand{\alth}{\ion{Al}{iii}} \newcommand{\few}{\ion{Fe}{ii}} 
\newcommand{\ssix}{\ion{S}{vi}}  \newcommand{\sfo}{\ion{S}{iv}}  
\newcommand{\lya}{Lyman-$\alpha$}
\newcommand{\zgrb}{$z_{\rm GRB}$}

\begin{document}

\title{High-ion absorption in seven GRB host galaxies at
  $z$=2--4\thanks{Based on observations taken under Programme IDs  
  070.A-0599, 070.D-0523, 075.A-0385, 075.A-0603, 077.D-0661, and
  080.D-0526 with the Ultraviolet and Visual
  Echelle Spectrograph (UVES) on the Very Large Telescope (VLT) Unit 2
  (Kueyen) at Paranal, Chile, operated by ESO.}}
\subtitle{Evidence for both circumburst plasma and outflowing
interstellar gas} 

\author{Andrew J. Fox\inst{1} \and C\'edric Ledoux\inst{1} 
 \and Paul M. Vreeswijk\inst{2} \and Alain Smette\inst{1}
 \and Andreas O. Jaunsen\inst{3}} 
\institute{European Southern Observatory, Alonso de C\'ordova
  3107, Casilla 19001, Vitacura, Santiago, Chile; afox@eso.org \and
  Dark Cosmology Centre, Niels Bohr Institute,
  U. of Copenhagen, Juliane Maries Vej 30, 2100 Copenhagen, Denmark 
 \and Institute of Theoretical Astrophysics, University of Oslo, PO
 Box 1029 Blindern, 0315 Oslo, Norway}

\date{Accepted 16 September 2008, Received 15 September 2008, in
  original form 29 May 2008}

\authorrunning{Fox et al.}
\titlerunning{High ions in GRB host galaxies} 

\abstract{}
{We use VLT/UVES high-resolution optical spectroscopy of
seven GRB afterglows (at \zgrb\ between 2.20 and 3.97) to investigate
circumburst and interstellar plasma in the host galaxies of the bursts.} 
{GRBs 050730, 050820, 050922C, 060607, 071031, and 080310 were each
detected by the \emph{Swift} satellite.
With the minimum possible time delay (as short as eight minutes), 
follow-up optical spectroscopy at 6.0~\kms\ resolution
began with UVES. We present Voigt-profile component fits and
analysis of the high-ion absorption detected in these spectra and in
the pre-\emph{Swift} UVES spectrum of GRB~021004, at velocities
between \zgrb--5\,000~\kms\ and \zgrb.} 
{
We identify several distinct categories of high-ion absorption at
velocities close to \zgrb:
(i) Strong high-ion absorption at \zgrb\ itself is always seen in
\os, \cf, and \sif, usually (in six of seven cases) in \nf,
and occasionally in \sfo\ and \ssix. 
Three of the cases show log\,$N$(\nf)$>$14 in a single-component,
suggesting a circumburst origin, but we cannot rule out an
interstellar origin. Indeed, using the 
non-detection of \sfo$^*$ at \zgrb\ toward GRB~050730 together with a UV
photo-excitation model, we place a lower limit of 
400~pc on the distance of the \sfo-bearing gas from the GRB.
(ii) Complex, multi-component \cf\ and \sif\ profiles 
extending over 100--400~\kms\ around \zgrb\ are observed 
in each spectrum; these velocity fields are similar to those measured
in \cf\ in damped \lya\ systems at similar redshifts, suggesting a
galactic origin.
(iii) Asymmetric, blueshifted, absorption-line wings covering
65--140~\kms\ are seen in the \cf, \sif, and \os\
profiles in four of the seven GRB afterglow spectra. 
The wing kinematics together with the ``Galactic'' \cf/\sif\ ratios
measured in two cases suggest that the wings
trace outflowing interstellar gas in the GRB host galaxies.  
(iv) High-velocity (HV; 500--5\,000~\kms\ relative to \zgrb) 
components are detected in six of the seven spectra; these components
are not necessarily a single homogeneous population, and many of them
may arise in unrelated foreground galaxies. However, in the cases of
GRBs~071031 and 080310, the ionization properties of the HV components
(very high \cf/\sif\ ratios and absence of neutral-phase absorption 
in \siw\ or \cw) are suggestive of a circumburst origin; models of
Wolf-Rayet winds from the GRB progenitors can explain both
the kinematics and ionization level of these HV components. 
}{}

\keywords{gamma rays: bursts - galaxies: halos - galaxies:
  high-redshift - galaxies: ISM - stars: Wolf-Rayet}

\maketitle
\section{Introduction}
For many years the only backlights available for studying the
distant Universe through absorption-line spectroscopy were AGN. 
Since the discovery that gamma-ray bursts (GRBs) are extragalactic 
\citep{vP97, Me97}, GRBs have also been employed as backlights, with the
advantages of extremely high luminosity and power-law continua, but the
disadvantage of rapid temporal fading, implying that rapid-response
observations are necessary for their study.
Not only do analyses of GRB afterglow spectra allow intervening systems to
be studied, they also enable the properties of the interstellar gas in
the host galaxy of the GRB to be investigated, particularly its
kinematics, chemical abundances, dust content, molecular gas content,
and ionized gas content 
\citep{Ca03, Kl04, Fi05, Sv03, Sv04, Pe06, Fy06, Pr06, Pr07a}.
Many GRB spectra show damped \lya\ (DLA) absorption at \zgrb, 
i.e. they show a neutral gas column density
log\,$N$(\hi)$>$20.3 \citep{Je01, Hj03,
  Ja04, Ja06, Vr04, St05, Be06, Wa06, Pr07b},  
typically with higher \hi\ column densities than are seen in DLAs
toward QSOs \citep[QSO-DLAs, reviewed by][]{Wo05}.
The higher $N$(\hi) in GRB-DLAs may be
related to their intense star-formation activity \citep{Bl02}, or to
the GRB sight-lines passing preferentially through the inner
regions of the host galaxies \citep{Fy08, Pr08a}.
Photometric studies of GRB host galaxies have shown them to be
star-forming, dwarf galaxies \citep{Ch04, Wi07, Th07, Th08a}, with no
evidence that the hosts are peculiar \citep{Sv08}. 

In this paper we use rapid-response mode spectra of seven
high-redshift long-duration GRB afterglows to investigate high-ion
absorption (in \os, \nf, \cf, \sif, \sfo, and \ssix) in the ISM of the
host galaxy, as well as in the circumburst medium immediately
surrounding the GRB. Observations of ionized interstellar gas in
galactic environments provide important constraints on several
physical processes, including star formation and subsequent feedback,
galactic winds, interactions with satellite galaxies, and accretion. 
In the pre-\emph{Swift} era, studying high ions in galaxy
environments at high redshifts was complicated by the difficulty of
associating individual absorbers with individual galaxies. 
There are several \emph{indirect} ways to explore high-ion absorption 
in high-redshift galaxies.
Firstly, outflowing \cf\ and \sif\ is seen in the composite spectrum of
high-redshift Lyman-break galaxies \citep[LBGs;][]{Sh03}.
Secondly, there is a statistical correlation between the redshifts of
\cf\ absorbers seen in QSO spectra and the redshifts of galaxies near
the QSO lines-of-sight \citep{Ad05}.
And thirdly, high-ion absorption is detected in all
DLAs \citep{WP00, Fo07a, Fo07b, Le08}, which trace galactic structures. 
However, because of the faintness of high-redshift galaxies,
only in a single published case \citep[the lensed LBG cB58;][]{Pe00, Pe02}
are interstellar high-ion absorption lines seen at moderate-to-high
resolution in the spectra of \emph{individual} galaxies at $z\!>\!2$.

Now, with the ability to be on-target within a few minutes of a
\emph{Swift} GRB trigger, a new means to directly probe the
interstellar (as well as circumburst) medium in individual high-redshift
galaxies is available \citep[see][]{Fi05, DE07, Ch07}.
High-resolution optical spectroscopy of GRB afterglows 
now routinely produces spectra with high-enough signal-to-noise to
uncover the velocity sub-structure in the absorbing gas. 
Evidence has even been found for time-variation in the strength of
absorption in fine-structure lines at \zgrb\ \citep{DZ06, Vr07}.  
In this paper, we present a survey of high-ion absorption in the
highest-quality GRB optical afterglow spectra taken to date.
We pay particular attention to the high-ion kinematics, since velocity
measurements can constrain the presence of gaseous outflows, either
from the GRB progenitor or from the host galaxy.
We structure this paper as follows.
In \S2 we discuss the GRB observations, data processing, and our line
profile fitting procedures. In \S3 we describe the observational
properties of the high-ion absorption at velocities near \zgrb\ in
each afterglow spectrum. We discuss four distinct observational
categories of high-ion absorption in the GRB afterglow spectra in \S4,
and we present a summary in \S5.

\section{GRB observations}
We formed our sample of seven GRB afterglow spectra by selecting 
all GRBs at $z\!>\!2$ with follow-up high-resolution
UVES\footnote{UVES is described in \citet{De00}.} optical spectroscopy
available as of March 2008. 
Lower redshift GRBs were not considered since no information on the
high ions (particularly \os) is available in these cases.
With the exception of GRB~021004, which was detected by the
\emph{High-Energy Transient Explorer (HETE-2)} satellite \citep{Ri03},
each of the GRBs in our sample was detected by 
the Burst Alert Telescope (BAT) on-board 
NASA's \emph{Swift} satellite \citep{Ge04}.
The seven GRBs are 021004, 050730, 050820, 050922C, 060607, 071031,
and 080310, named according to the UT date of their detection (yymmdd). 
Following each trigger, \emph{Swift} began follow-up
observations with the XRT and UVOT instruments, to determine an
error circle of $\approx$2.5\arcsec\ around the source.  

After the \emph{Swift} localization was distributed through the GCN 
(GRB Coordinates Network), rapid-response mode (RRM) UVES observations
began on each GRB afterglow with the minimum possible time delay,
provided the target was observable from Paranal. Three of the seven GRB
afterglows in our sample were observed in this mode.
RRM observations involve the immediate interruption of any current UVES
exposure in order to slew to and acquire the new target. No human
intervention is required, except for alignment of the target on the
spectrograph slit. On two occasions, a sub-ten minute response was 
achieved. These observations demonstrate the power of the RRM concept,
and the success of its implementation on ESO telescopes. 
The four afterglow spectra that were not taken in the RRM mode are 
GRBs~021004, 050820, 050730 and 050922C, which either occurred
before the RRM was implemented or were unobservable from Paranal at
the time of the trigger. In each of these cases, the UVES 
observations began at the earliest possible time in the form of a
traditional Target-of-Opportunity (ToO) observation.
A log of the observations, together with a summary of
the basic properties of the GRBs, is given in Table 1.

The spectra were taken with 2$\times$2 binning and a 1.0\arcsec\ slit,
providing $R\approx50\,000$ (FWHM=6.0~\kms) spectra in
$\approx$2~\kms\ pixels.\footnote{Though the nominal spectral resolution of
  UVES in this mode is 43\,000, we find that a higher resolution of
  $\approx$50\,000 is achieved in practice, due to variations in the seeing
  conditions.} 
The UVES data were reduced with a customized version of the {\sc MIDAS}
reduction pipeline \citep{Ba00}, with the wavelength scale
subsequently corrected into the vacuum heliocentric frame. 
For the latest GRB in our sample (GRB~080310), we also reduced the
data independently with version 3.4.5 of the CPL (Common Pipeline Library) 
pipeline\footnote{See manual at 
ftp://ftp.eso.org/pub/dfs/pipelines/uves/.}.
A detailed comparison of line profiles found the results from the two
pipeline versions to be essentially identical.
Individual exposures were combined with a least-squares weighting 
to produce a combined spectrum. 
In the case of GRB~071031, we also maintain and analyze the
individual exposures to look for time-series variation (see \S3.6.1). 
Continua were fit locally to each absorption line of interest.
Before displaying and fitting the \os, \ct, \nf, and \ssix\ profiles,
which fall in low S/N regions of the spectrum, we perform a further
rebinning, generally by five pixels, but occasionally by other
factors, as described in the captions to Figures 1 to 4. These
differing rebinning factors are driven by the desire to maximize
the signal-to-noise ratio and to investigate the significance of the
high-ion detections, even if some resolution is compromised.

\begin{table*}
\begin{minipage}[t]{18cm}
\caption{GRB sample and log of UVES observations}
\begin{tabular}{lcccc ccccc c}
\hline\hline
GRB & UT(Trig)\footnote{UT of trigger by the BAT instrument on-board
  \emph{Swift}. Exception: GRB~020104, detected by \emph{WXM} 
  on-board \emph{HETE-2}.} & UT(UVES)\footnote{UT of start of first UVES
  exposure (after target acquisition).} &  
$\delta t$\footnote{Time delay between satellite trigger and start
  of UVES exposure.} & $t_{\rm total}$\footnote{Total UVES exposure
  time over all setups.} & ESO ID & \zgrb\ & log\,$N_{\rm
  \hi}$\footnote{\hi\ column density and metallicity of GRB-DLA taken from
  reference 8, except 071031 and 080310 values, from this 
  paper.} & [Z/H]$^e$ & GCN\footnote{We list here the GCN first
  reporting the burst, and the GCN first reporting the UVES spectrum.} 
  & Science\footnote{We list references relevant to the study of the
  host galaxy ISM.}\\
(yymmdd) & & (hh:mm) & (hh:mm) & (s) & & & & & Refs. & Refs.\\ 
\hline
021004  & 12:06:13 & 01:37 & 13:31 &  7200 & 070.A-0599\footnote{Also 070.D-0523.} 
                                                           & 2.3290 & 19.00 &    0.0 & 1,2   & 3,4,5,6,7,8\\
050730  & 19:58:23 & 00:07 & 04:09 &  6000 & 075.A-0603 & 3.9686 & 22.15 & $-$2.3 & 9,10  & 8,11,12,13,14,15,16\\
050820  & 06:34:53 & 07:08 & 00:34 &  6043 & 075.A-0385 & 2.6147 & 21.00 & $-$0.6 & 17,18 & 8,13,15,16\\
050922C & 19:55:50 & 23:42 & 03:47 &  6000 & 075.A-0603 & 2.1990 & 21.55 & $-$2.0 & 19,20 & 8,16,21\\
060607  & 05:12:13 & 05:20 & 00:08 & 11980 & 077.D-0661 & 3.0749 & 16.80 &    0.0 & 22,23 & 8\\
071031  & 01:06:36 & 01:15\footnote{Due to a difficult target acquisition,
  the first fully aligned exposure of GRB~071031 began at 01:25.}& 00:09 &  9480 & 080.D-0526 & 2.6922 & 22.15 & $-$1.7 & 24,25 & ...\\
080310  & 08:37:58 & 08:51 & 00:13 &  4680 & 080.D-0526 & 2.4274 & 18.80 & $-$1.4 & 26,27 & ...\\
\hline
\end{tabular} 
\end{minipage}
REFERENCES: 1 \citet{Sh02}; 2 \citet{Sv02}; 
3 \citet{Mo02}; 4 \citet{Mi03}; 5 \citet{Sc03}; 6 \citet{Fi05}; 7 \citet{La06}; 8 \citet{Pr08b}; 
9 \citet{Ho05}; 10 \citet{DE05a}; 
11 \citet{St05}; 12 \citet{Ch05}; 13 \citet{Ch07}; 14 \citet{DE07}; 15 \citet{Pr07a}; 16 \citet{Pr07b}; 
17 \citet{Pa05}; 18 \citet{Le05}; 
19 \citet{No05}; 20 \citet{DE05b}; 21 \citet{Ja06}; 
22 \citet{Zi06}; 23 \citet{Le06};
24 \citet{St07}; 25 \citet{Le07}; 
26 \citet{Cu08}; 27 \citet{Vr08}.
\end{table*}

We searched in each spectrum for high-ion absorption over the velocity
range \zgrb--5\,000~\kms\ to \zgrb+2\,000~\kms, where \zgrb\ is defined by
the position of strongest absorption in the low-ionization lines.
This wide range was chosen because of the possibility of detecting
high-velocity outflow features driven by the burst progenitor.
We make a distinction between low-velocity (LV) absorption components
seen within several hundred~\kms\ of \zgrb, which are seen in all cases, and
high-velocity (HV) components at 500-5\,000~\kms\ from \zgrb, which
are seen in 6/7 sight lines, and which are clearly separated from the LV
absorption. When absorption was found, we used the
VPFIT software package\footnote{Available at 
http://www.ast.cam.ac.uk/$\sim$rfc/vpfit.html.} to fit the high-ion
components in each GRB spectrum with a series of Voigt profiles. 
Each high ion was fit independently, even species with similar ionization
potentials. The VPFIT software accounts for instrumental resolution, and
returns the redshift, line width, and column density ($z$, $b$, and
log\,$N$) for each component together with the associated errors.
We convert the component redshift to velocity, using \zgrb\ as the
zero point. By summing over the component column densities, and adding
the component errors in quadrature, we form the total column density
and its error in each absorber. 
In cases where the high-ion absorption is saturated, the
errors on the total column density are large.
The fit solutions are non-unique, since the number of components to be
fit has to be specified manually. This number is often larger than
ten, due to the complexity of the absorption-line profiles. 
Nonetheless, our fitting technique has been tested by comparing the
total model column densities with the values from direct apparent
optical depth integrations \citep{Fo07a}. The successful outcome of
these tests supports the reliability of the fit results.  
The details of the fits are given in Tables 2 to 8, for 
GRBs 021004, 050730, 050820, 050922C, 060607, 071031, and 080310,
respectively. Atomic data (rest wavelengths and $f$-values)
were taken from \citet{Mo03}.

\begin{table}
\begin{minipage}[t]{\columnwidth}
\caption{GRB 021004 $z$=2.3290: Voigt profile fits}
\begin{tabular}{lcccc}
\hline\hline
Ion & $v_0$ & $b$ & log\,$N$ & log\,$N$(total)\\
 & (\kms) & (\kms) & ($N$ in \sqcm) & ($N$ in \sqcm)\\
\hline
       \os & $-$25$\pm$8 &  18$\pm$10 & 14.46$\pm$0.22 &            14.95$\pm$0.20 \\
       ... &    26$\pm$7 &  22$\pm$8 & 14.78$\pm$0.22 &                       ... \\
       \nf & $-$2903$\pm$2 &  19$\pm$2 & 14.03$\pm$0.05 &            14.03$\pm$0.05 \\
       \nf & $-$10$\pm$2 &  28$\pm$3 & 14.53$\pm$0.06 &            14.53$\pm$0.06 \\
       \cf & $-$2966$\pm$4 &  12$\pm$6 & 12.97$\pm$0.19 &            14.73$\pm$0.08 \\
       ... & $-$2903$\pm$2 &  30$\pm$2 & 14.59$\pm$0.04 &                       ... \\
       ... & $-$2720$\pm$2 &   8$\pm$2 & 13.36$\pm$0.09 &                       ... \\
       ... & $-$2670$\pm$2 &  31$\pm$3 & 14.03$\pm$0.04 &                       ... \\
       ... & $-$2599$\pm$8 &  24$\pm$11 & 13.10$\pm$0.18 &                       ... \\
      \sif & $-$2903$\pm$3    & $^*$ & $^*$ &             $\approx$14.82$^*$ \\
       ... & $-$2882$\pm$45 &  58$\pm$40 & 13.53$\pm$0.42 &                       ... \\
       ... & $-$2659$\pm$4 &  20$\pm$5 & 13.20$\pm$0.10 &                       ... \\
      \sfo &     0$\pm$4 &  45$\pm$5 & 15.25$\pm$0.20 &            15.25$\pm$0.20 \\
\hline
\end{tabular}
\end{minipage}
NOTE: the symbol $^*$ denotes a saturated component; no reliable column density or line width can be extracted for the component, and the total column density in that ion is uncertain.
NOTE: Total column densities (summed over components) are reported separately for the HV and LV absorption.
NOTE: No \cf\ or \sif\ fit was attempted to the absorption within 1000~\kms\ of \zgrb, due to complete saturation.
\end{table}

\begin{table}
\begin{minipage}[t]{\columnwidth}
\caption{GRB 050730 $z$=3.9686: Voigt profile fits}
\begin{tabular}{lcccc}
\hline\hline
Ion & $v_0$ & $b$ & log\,$N$ & log\,$N$(total)\\
 & (\kms) & (\kms) & ($N$ in \sqcm) & ($N$ in \sqcm)\\
\hline
       \os & $-$88$\pm$5 &  34$\pm$5 & 14.52$\pm$0.08 &        $\approx$15.94$^*$ \\
       ... & $-$54$\pm$2    & $^*$ & $^*$ &                            ... \\
       ... &     1$\pm$8 &  36$\pm$6 & 14.97$\pm$0.15 &                       ... \\
       ... &    31$\pm$6    & $^*$ & $^*$ &                            ... \\
       \nf & $-$ 1$\pm$2 &   7$\pm$3 & 13.54$\pm$0.15 &            13.80$\pm$0.14 \\
       ... &    26$\pm$5 &  14$\pm$8 & 13.46$\pm$0.18 &                       ... \\
       \cf & $-$1577$\pm$2 &   4$\pm$2 & 13.38$\pm$0.22 &            14.52$\pm$0.18 \\
       ... & $-$1567$\pm$2 &   5$\pm$4 & 13.43$\pm$0.23 &                       ... \\
       ... & $-$1553$\pm$2 &   5$\pm$2 & 13.66$\pm$0.17 &                       ... \\
       ... & $-$1536$\pm$2 &  12$\pm$2 & 13.97$\pm$0.10 &                       ... \\
       ... & $-$1516$\pm$2 &  34$\pm$34 & 13.81$\pm$0.37 &                       ... \\
       ... & $-$1480$\pm$2 &  11$\pm$3 & 13.86$\pm$0.17 &                       ... \\
       \cf & $-$168$\pm$2 &  14$\pm$3 & 13.12$\pm$0.06 &        $\approx$15.24$^*$ \\
       ... & $-$119$\pm$18    & $^*$ & $^*$ &                            ... \\
       ... & $-$98$\pm$7    & $^*$ & $^*$ &                            ... \\
       ... & $-$58$\pm$2 &  17$\pm$3 & 14.81$\pm$0.10 &                       ... \\
       ... &     4$\pm$2 &  21$\pm$2 & 15.01$\pm$0.06 &                       ... \\
      \sif & $-$1575$\pm$2 &   4$\pm$2 & 12.57$\pm$0.06 &            13.94$\pm$0.10 \\
       ... & $-$1553$\pm$2 &   4$\pm$2 & 13.03$\pm$0.05 &                       ... \\
       ... & $-$1542$\pm$3 &  26$\pm$3 & 13.38$\pm$0.08 &                       ... \\
       ... & $-$1529$\pm$2 &   7$\pm$4 & 12.45$\pm$0.38 &                       ... \\
       ... & $-$1499$\pm$2 &   9$\pm$4 & 12.61$\pm$0.21 &                       ... \\
       ... & $-$1472$\pm$2 &  12$\pm$2 & 13.50$\pm$0.03 &                       ... \\
       ... & $-$1442$\pm$2 &  13$\pm$2 & 13.02$\pm$0.04 &                       ... \\
      \sif & $-$104$\pm$6    & $^*$ & $^*$ &             $\approx$14.35$^*$ \\
       ... & $-$63$\pm$50    & $^*$ & $^*$ &                            ... \\
       ... & $-$35$\pm$50    & $^*$ & $^*$ &                            ... \\
       ... &     1$\pm$2 &  10$\pm$2 & 13.94$\pm$0.12 &                       ... \\
      \sfo & $-$53$\pm$2 &  28$\pm$4 & 14.51$\pm$0.06 &        $\approx$15.31$^*$ \\
       ... & $-$55$\pm$2    & $^*$ & $^*$ &                            ... \\
       ... &     1$\pm$2 &   8$\pm$2 & 14.65$\pm$0.06 &                       ... \\
       ... &    21$\pm$2 &   4$\pm$3 & 13.57$\pm$0.15 &                       ... \\
     \ssix & $-$ 3$\pm$2 &  17$\pm$2 & 14.16$\pm$0.06 &            14.16$\pm$0.06 \\
\hline
\end{tabular}
\end{minipage}
NOTE: the symbol $^*$ denotes a saturated component; no reliable column density or line width can be extracted for the component, and the total column density in that ion is uncertain.
NOTE: Total column densities (summed over components) are reported separately for the HV and LV absorption.
\end{table}

\begin{table}
\begin{minipage}[t]{\columnwidth}
\caption{GRB 050820 $z$=2.6147: Voigt profile fits}
\begin{tabular}{lcccc}
\hline\hline
Ion & $v_0$ & $b$ & log\,$N$ & log\,$N$(total)\\
 & (\kms) & (\kms) & ($N$ in \sqcm) & ($N$ in \sqcm)\\
\hline
       \nf & $-$109$\pm$2 &   9$\pm$3 & 13.17$\pm$0.08 &            13.17$\pm$0.08 \\
       \cf & $-$3655$\pm$2 &  10$\pm$2 & 13.02$\pm$0.04 &            13.13$\pm$0.06 \\
       ... & $-$3631$\pm$2 &  10$\pm$3 & 12.50$\pm$0.11 &                       ... \\
       \cf & $-$268$\pm$6 &  65$\pm$8 & 13.42$\pm$0.05 &            14.91$\pm$0.11 \\
       ... & $-$247$\pm$2 &   9$\pm$2 & 13.87$\pm$0.03 &                       ... \\
       ... & $-$223$\pm$2 &   9$\pm$2 & 13.53$\pm$0.03 &                       ... \\
       ... & $-$167$\pm$2 &   9$\pm$2 & 13.13$\pm$0.07 &                       ... \\
       ... & $-$146$\pm$2 &   9$\pm$2 & 13.58$\pm$0.04 &                       ... \\
       ... & $-$109$\pm$2 &  18$\pm$2 & 14.02$\pm$0.02 &                       ... \\
       ... & $-$62$\pm$2 &  15$\pm$2 & 13.74$\pm$0.06 &                       ... \\
       ... & $-$15$\pm$2 &  25$\pm$2 & 14.55$\pm$0.04 &                       ... \\
       ... &    84$\pm$2 &  35$\pm$2 & 14.06$\pm$0.02 &                       ... \\
      \sif & $-$249$\pm$2 &   5$\pm$2 & 12.23$\pm$0.05 &            14.35$\pm$0.18 \\
       ... & $-$222$\pm$2 &   8$\pm$3 & 11.98$\pm$0.10 &                       ... \\
       ... & $-$170$\pm$2 &   8$\pm$2 & 12.72$\pm$0.04 &                       ... \\
       ... & $-$145$\pm$2 &  10$\pm$2 & 13.05$\pm$0.06 &                       ... \\
       ... & $-$122$\pm$3 &  13$\pm$4 & 12.67$\pm$0.14 &                       ... \\
       ... & $-$57$\pm$2 &   7$\pm$2 & 12.74$\pm$0.14 &                       ... \\
       ... & $-$37$\pm$3 &  46$\pm$3 & 13.72$\pm$0.08 &                       ... \\
       ... & $-$12$\pm$2 &  20$\pm$2 & 14.00$\pm$0.04 &                       ... \\
       ... &    64$\pm$2 &  16$\pm$2 & 12.96$\pm$0.06 &                       ... \\
       ... &    91$\pm$2 &  11$\pm$3 & 12.82$\pm$0.09 &                       ... \\
       ... &    111$\pm$2 &   4$\pm$3 & 12.14$\pm$0.20 &                       ... \\
       ... &    147$\pm$2 &  26$\pm$3 & 13.23$\pm$0.05 &                       ... \\
       ... &    168$\pm$2 &   4$\pm$3 & 12.05$\pm$0.19 &                       ... \\
       ... &    220$\pm$2 &   4$\pm$2 & 12.39$\pm$0.13 &                       ... \\
       ... &    202$\pm$6 &  17$\pm$12 & 12.32$\pm$0.29 &                       ... \\
      \sfo & $-$145$\pm$2 &  25$\pm$3 & 14.04$\pm$0.04 &            14.91$\pm$0.07 \\
       ... & $-$67$\pm$2 &   4$\pm$3 & 13.30$\pm$0.10 &                       ... \\
       ... & $-$ 5$\pm$2 &   4$\pm$2 & 14.04$\pm$0.12 &                       ... \\
       ... & $-$ 8$\pm$2 &  32$\pm$2 & 14.76$\pm$0.03 &                       ... \\
     \ssix & $-$109$\pm$3 &  20$\pm$4 & 13.45$\pm$0.07 &            14.09$\pm$0.08 \\
       ... & $-$49$\pm$3 &  14$\pm$3 & 13.54$\pm$0.11 &                       ... \\
       ... & $-$11$\pm$3 &  17$\pm$3 & 13.77$\pm$0.07 &                       ... \\
\hline
\end{tabular}
\end{minipage}
NOTE: Total column densities (summed over components) are reported separately for the HV and LV absorption.
NOTE: \ssix\ is only detected in $\lambda$933 ($\lambda$944 is blended), so the detection is uncertain. Nonetheless the velocity structure corresponds closely to \cf.
\end{table}

\begin{table}
\begin{minipage}[t]{\columnwidth}
\caption{GRB 050922C $z$=2.1990: Voigt profile fits}
\begin{tabular}{lcccc}
\hline\hline
Ion & $v_0$ & $b$ & log\,$N$ & log\,$N$(total)\\
 & (\kms) & (\kms) & ($N$ in \sqcm) & ($N$ in \sqcm)\\
\hline
       \os &    65$\pm$2 &  24$\pm$2 & 14.44$\pm$0.05 &            14.44$\pm$0.05 \\
       \nf &    63$\pm$2 &  16$\pm$3 & 13.76$\pm$0.07 &            13.76$\pm$0.07 \\
       \cf & $-$20$\pm$2 &  16$\pm$2 & 13.39$\pm$0.03 &            15.69$\pm$0.35 \\
       ... &    25$\pm$2 &  10$\pm$2 & 13.21$\pm$0.08 &                       ... \\
       ... &    63$\pm$2 &  11$\pm$2 & 15.67$\pm$0.30 &                       ... \\
       ... &    96$\pm$2 &   8$\pm$2 & 13.63$\pm$0.07 &                       ... \\
       ... &    140$\pm$2 &  18$\pm$2 & 13.73$\pm$0.03 &                       ... \\
       ... &    201$\pm$2 &  10$\pm$2 & 13.18$\pm$0.04 &                       ... \\
      \sif &    27$\pm$2 &   7$\pm$2 & 13.22$\pm$0.04 &            14.49$\pm$0.16 \\
       ... &    62$\pm$2 &  12$\pm$2 & 14.40$\pm$0.11 &                       ... \\
       ... &    85$\pm$2 &   5$\pm$3 & 13.13$\pm$0.24 &                       ... \\
       ... &    98$\pm$2 &   6$\pm$2 & 13.28$\pm$0.05 &                       ... \\
       ... &    127$\pm$2 &   4$\pm$2 & 12.42$\pm$0.07 &                       ... \\
       ... &    141$\pm$2 &   4$\pm$2 & 12.70$\pm$0.05 &                       ... \\
       ... &    198$\pm$2 &   7$\pm$2 & 12.53$\pm$0.07 &                       ... \\
\hline
\end{tabular}
\end{minipage}
\end{table}

\begin{table}
\begin{minipage}[t]{\columnwidth}
\caption{GRB 060607 $z$=3.0749: Voigt profile fits}
\begin{tabular}{lcccc}
\hline\hline
Ion & $v_0$ & $b$ & log\,$N$ & log\,$N$(total)\\
 & (\kms) & (\kms) & ($N$ in \sqcm) & ($N$ in \sqcm)\\
\hline
       \cf & $-$1849$\pm$2 &   9$\pm$2 & 13.70$\pm$0.03 &            13.83$\pm$0.05 \\
       ... & $-$1824$\pm$2 &  18$\pm$2 & 13.23$\pm$0.05 &                       ... \\
       \cf & $-$88$\pm$3 &  31$\pm$4 & 13.00$\pm$0.05 &            15.94$\pm$0.32 \\
       ... & $-$45$\pm$2 &  10$\pm$2 & 13.03$\pm$0.06 &                       ... \\
       ... & $-$ 5$\pm$2 &  19$\pm$2 & 14.26$\pm$0.02 &                       ... \\
       ... &    19$\pm$2 &   4$\pm$2 & 15.93$\pm$0.28 &                       ... \\
      \sif & $-$1848$\pm$2 &  11$\pm$2 & 13.12$\pm$0.04 &            13.28$\pm$0.06 \\
       ... & $-$1817$\pm$2 &  12$\pm$3 & 12.76$\pm$0.07 &                       ... \\
      \sif & $-$47$\pm$2 &   4$\pm$2 & 12.31$\pm$0.12 &            13.54$\pm$0.09 \\
       ... & $-$30$\pm$3 &   8$\pm$5 & 12.48$\pm$0.23 &                       ... \\
       ... & $-$ 6$\pm$2 &  12$\pm$2 & 13.30$\pm$0.06 &                       ... \\
       ... &    16$\pm$2 &  10$\pm$2 & 12.98$\pm$0.07 &                       ... \\
\hline
\end{tabular}
\end{minipage}
NOTE: Total column densities (summed over components) are reported separately for the HV and LV absorption.
\end{table}

\begin{table}
\begin{minipage}[t]{\columnwidth}
\caption{GRB 071031 $z$=2.6922: Voigt profile fits}
\begin{tabular}{lcccc}
\hline\hline
Ion & $v_0$ & $b$ & log\,$N$ & log\,$N$(total)\\
 & (\kms) & (\kms) & ($N$ in \sqcm) & ($N$ in \sqcm)\\
\hline
       \os & $-$560$\pm$3 &  18$\pm$3 & 14.52$\pm$0.08 &            14.77$\pm$0.08 \\
       ... & $-$505$\pm$2 &  14$\pm$2 & 14.41$\pm$0.06 &                       ... \\
       \os & $-$118$\pm$7 &  13$\pm$12 & 13.79$\pm$0.42 &        $\approx$15.14$^*$ \\
       ... & $-$81$\pm$15    & $^*$ & $^*$ &                            ... \\
       ... & $-$31$\pm$12    & $^*$ & $^*$ &                            ... \\
       ... &     6$\pm$14    & $^*$ & $^*$ &                            ... \\
       ... &    31$\pm$15    & $^*$ & $^*$ &                            ... \\
       \nf &    25$\pm$2 &  11$\pm$2 & 14.41$\pm$0.15 &            14.41$\pm$0.15 \\
       \cf & $-$566$\pm$2    & $^*$ & $^*$ &             $\approx$14.40$^*$ \\
       ... & $-$559$\pm$5 &  10$\pm$3 & 13.06$\pm$0.27 &                       ... \\
       ... & $-$515$\pm$3    & $^*$ & $^*$ &                            ... \\
       ... & $-$505$\pm$4 &   7$\pm$3 & 13.67$\pm$0.24 &                       ... \\
       ... & $-$368$\pm$2 &  12$\pm$2 & 13.93$\pm$0.03 &                       ... \\
       ... & $-$325$\pm$2 &  16$\pm$3 & 13.14$\pm$0.05 &                       ... \\
       \cf & $-$212$\pm$2 &   4$\pm$3 & 12.47$\pm$0.11 &        $\approx$15.11$^*$ \\
       ... & $-$147$\pm$2 &  11$\pm$2 & 13.13$\pm$0.06 &                       ... \\
       ... & $-$111$\pm$2 &  14$\pm$2 & 13.96$\pm$0.08 &                       ... \\
       ... & $-$70$\pm$6    & $^*$ & $^*$ &                            ... \\
       ... & $-$35$\pm$14 &  33$\pm$30 & 14.43$\pm$0.44 &                       ... \\
       ... &     8$\pm$7 &  22$\pm$5 & 14.69$\pm$0.22 &                       ... \\
       ... &    39$\pm$5 &   8$\pm$3 & 14.58$\pm$0.36 &                       ... \\
      \sif & $-$367$\pm$2 &   6$\pm$2 & 12.33$\pm$0.08 &            12.33$\pm$0.08 \\
      \sif & $-$109$\pm$2 &  15$\pm$2 & 13.28$\pm$0.03 &            15.93$\pm$0.51 \\
       ... & $-$63$\pm$2 &  19$\pm$3 & 13.61$\pm$0.07 &                       ... \\
       ... & $-$40$\pm$2 &   4$\pm$16 & 15.89$\pm$0.48 &                       ... \\
       ... & $-$ 1$\pm$3 &  29$\pm$3 & 14.22$\pm$0.07 &                       ... \\
       ... &    29$\pm$3 &  11$\pm$2 & 14.65$\pm$0.24 &                       ... \\
       \ct & $-$560$\pm$2 &  15$\pm$3 & 13.20$\pm$0.07 &        $\approx$14.47$^*$ \\
       ... & $-$507$\pm$2    & $^*$ & $^*$ &                            ... \\
       ... & $-$367$\pm$2    & $^*$ & $^*$ &                            ... \\
\hline
\end{tabular}
\end{minipage}
NOTE: the symbol $^*$ denotes a saturated component; no reliable column density or line width can be extracted for the component, and the total column density in that ion is uncertain.
NOTE: Total column densities (summed over components) are reported separately for the HV and LV absorption.
NOTE: \os\ is only detected at $-$118\,\kms\ in $\lambda$1031 ($\lambda$1037 is blended), so the detection in this component is uncertain.
\end{table}

\begin{table}
\begin{minipage}[t]{\columnwidth}
\caption{GRB 080310 $z$=2.4274: Voigt profile fits}
\begin{tabular}{lcccc}
\hline\hline
Ion & $v_0$ & $b$ & log\,$N$ & log\,$N$(total)\\
 & (\kms) & (\kms) & ($N$ in \sqcm) & ($N$ in \sqcm)\\
\hline
       \os & $-$137$\pm$4 &  25$\pm$5 & 14.18$\pm$0.08 &            15.12$\pm$0.09 \\
       ... & $-$30$\pm$3 &  42$\pm$4 & 15.07$\pm$0.08 &                       ... \\
       \nf & $-$49$\pm$2 &  18$\pm$3 & 13.75$\pm$0.05 &            14.03$\pm$0.05 \\
       ... &     1$\pm$2 &  19$\pm$3 & 13.71$\pm$0.05 &                       ... \\
       \cf & $-$1414$\pm$2 &   4$\pm$2 & 12.89$\pm$0.10 &            13.45$\pm$0.08 \\
       ... & $-$1388$\pm$4 &  33$\pm$5 & 13.31$\pm$0.07 &                       ... \\
       \cf & $-$222$\pm$16 &  22$\pm$14 & 13.03$\pm$0.41 &        $\approx$16.66$^*$ \\
       ... & $-$187$\pm$4 &  17$\pm$10 & 13.36$\pm$0.31 &                       ... \\
       ... & $-$160$\pm$3 &  11$\pm$4 & 13.33$\pm$0.25 &                       ... \\
       ... & $-$138$\pm$2 &   5$\pm$2 & 13.44$\pm$0.36 &                       ... \\
       ... & $-$66$\pm$50    & $^*$ & $^*$ &                            ... \\
       ... & $-$10$\pm$50    & $^*$ & $^*$ &                            ... \\
       ... &    69$\pm$2 &   5$\pm$2 & 12.88$\pm$0.08 &                       ... \\
      \sif & $-$136$\pm$2 &   4$\pm$2 & 12.53$\pm$0.05 &        $\approx$15.00$^*$ \\
       ... & $-$55$\pm$50    & $^*$ & $^*$ &                            ... \\
       ... & $-$29$\pm$19    & $^*$ & $^*$ &                            ... \\
       ... &     9$\pm$18    & $^*$ & $^*$ &                            ... \\
\hline
\end{tabular}
\end{minipage}
NOTE: the symbol $^*$ denotes a saturated component; no reliable column density or line width can be extracted for the component, and the total column density in that ion is uncertain.
NOTE: Total column densities (summed over components) are reported separately for the HV and LV absorption.
\end{table}

\section{Description of the high-ion absorption}
High-ion absorption within several hundred \kms\ of \zgrb\ is found in 
all seven GRB spectra in our sample.
The detected lines are \cf\ and \sif\ (seen in 7/7 cases),
\nf\ (6/7), \os\ (6/7), \sfo\ (5/7), and \ssix\ (2/7).
The high-ion absorption-line profiles in the velocity range within 
200~\kms\ of \zgrb\ are shown for each of our seven GRB afterglow
spectra in Figures 1 to 4. In addition to the high ions, we include on
each figure absorption lines that trace the neutral gas, typically 
\siw\ $\lambda$1260.422, \cw\ $\lambda$1334.532, or
\ion{Fe}{ii} $\lambda$1608.451. 

In cases where we do not report a detection for \os, \sfo, and \ssix\
(e.g. \os\ toward GRB~060607), it is because the profiles are blended, not
because there are actual non-detections in these ions.
The only clear non-detection at \zgrb\ in any of the high ions mentioned above
is in \nf\ toward GRB~060607. In two cases our data cover the wavelength of
\ion{Ne}{viii} $\lambda\lambda$770.409, 780.324 at \zgrb, but 
the doublet lies below the Lyman limit where no flux is transmitted.
\ion{P}{v} $\lambda\lambda$1117.977, 1128.008 is not detected in any of
the seven spectra. \ct\ $\lambda$977.020 is detected in one case
(GRB~071031), but in the other sight lines the S/N is too low at the
observed wavelength (far into the blue) for the \ct\ data to be useful. 
Absorption in \sfo$^*$~$\lambda$1072.973 is not observed in any of the
seven spectra. Finally, with the exception of GRB~060607 with its
unusually low $N$(\hi), \sit\ $\lambda$1206.500 lies in the strong
damping wings of \lya, so no useful information on this ion at \zgrb\
is available.

\begin{figure}
\includegraphics[width=8.0cm]{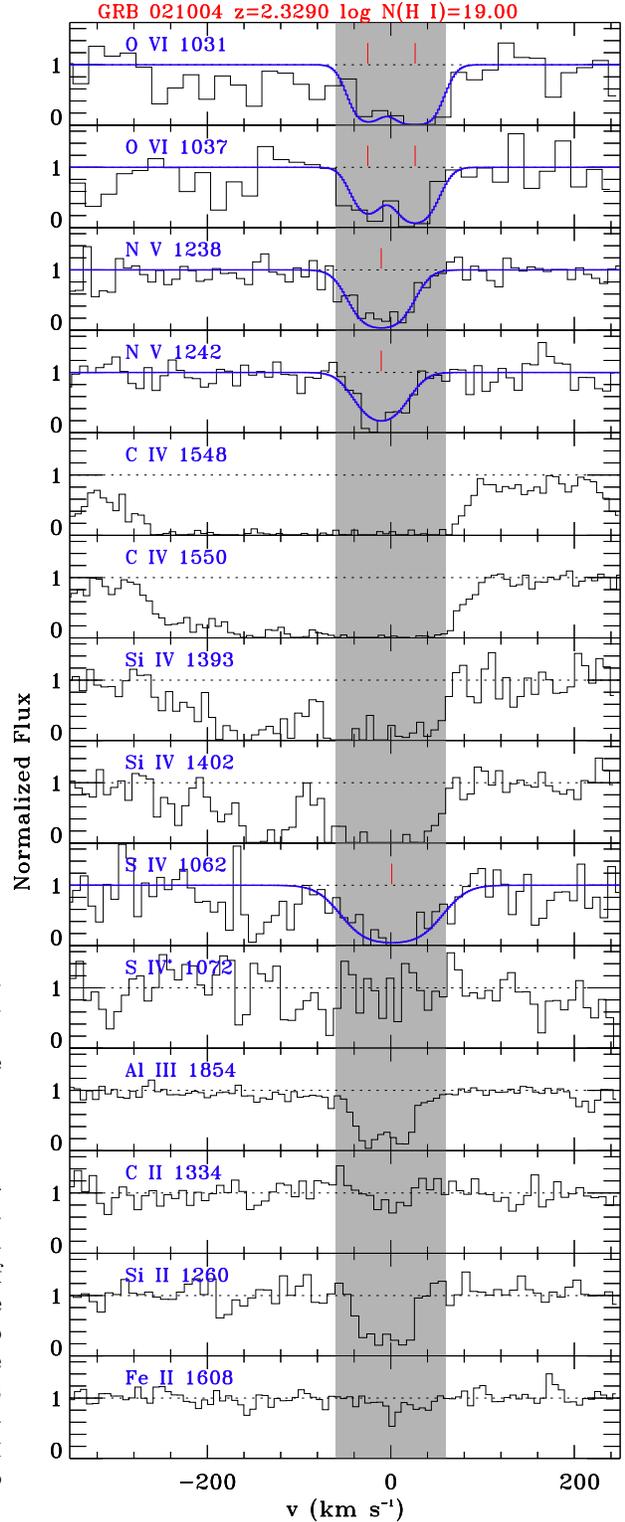}
\caption{Normalized high-ion and low-ion absorption-line profiles 
  on a velocity scale relative to \zgrb\ for GRB~021004.
  Voigt profile fits are shown in blue (solid line), with the center of each
  component in the fit marked with a red dash.
  Dark shading denotes the strong absorption components at \zgrb.
  All data have been rebinned by four pixels (eight in the case of
  \os). We do not fit the \cf\ or
  \sif\ doublets because of saturation. Absorption components at higher
  velocity relative to \zgrb\ are shown in Figure 5.}
\end{figure}

\begin{figure*}
\includegraphics[width=18cm]{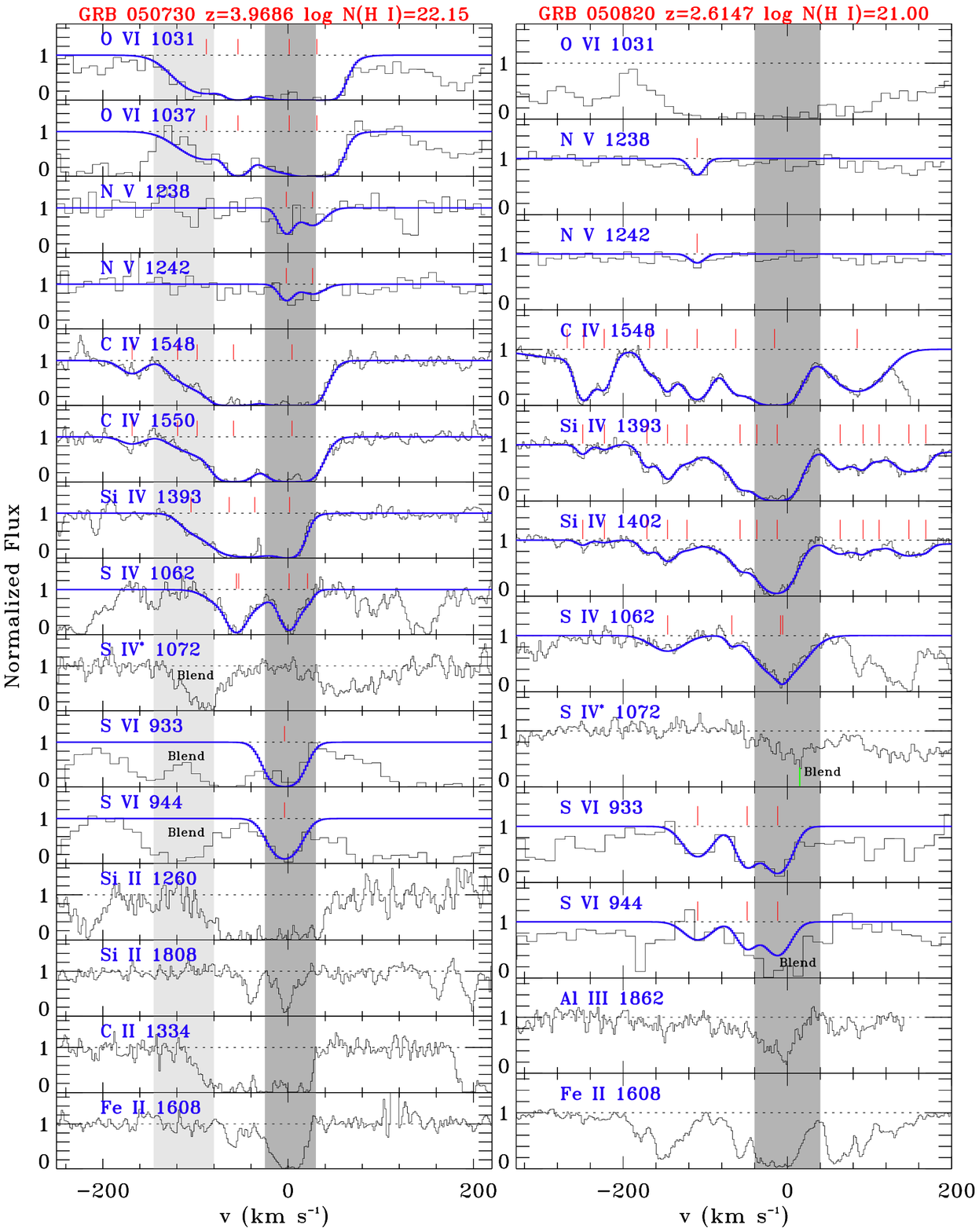}
\caption{Normalized high-ion and low-ion absorption-line profiles on a velocity
  scale relative to \zgrb\ for GRBs 050730 and 050820.  
  Voigt profile fits are shown in blue (solid line), with the center of each
  component in the fit marked with a red dash.
  Dark shading denotes the regions of strong absorption at \zgrb\
  itself, and light shading denotes the negative-velocity wings
  (GRB~050730 only). The \os, \nf, and \ssix\ data have been
  rebinned by five pixels. Absorption components at higher
  velocity relative to \zgrb\ are shown in Figure 5.}
\end{figure*}

\begin{figure*}
\includegraphics[width=18cm]{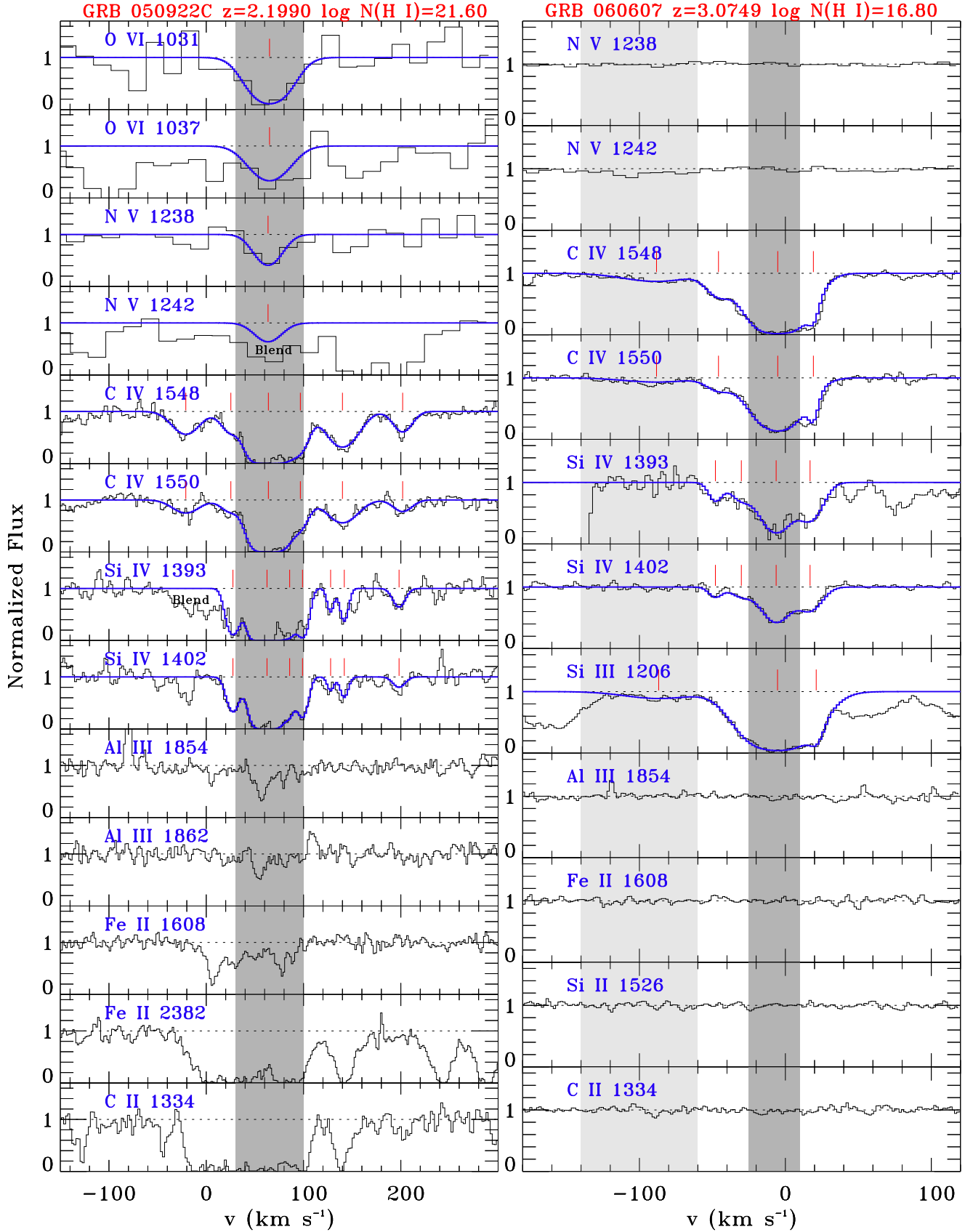}
\caption{Normalized high-ion and low-ion absorption-line profiles 
  on a velocity scale relative to \zgrb\ for GRBs 050922C and 060607. 
  Voigt profile fits are shown in blue (solid line), with the center of each
  component in the fit marked with a red dash.
  Dark shading denotes the regions of strong absorption,
  and light shading denotes the negative-velocity wing 
  (GRB~060607 only). 
  The \sfo\ and \ssix\ profiles are blended in each of these sight
  lines. The \os\ and \nf\ data have been rebinned
  by eight pixels for GRB~050922C, and the \nf\ data by five pixels
  for GRB~060607. Absorption components at higher
  velocity relative to \zgrb\ are shown in Figure 5.}
\end{figure*}

\begin{figure*}
\includegraphics[width=18cm]{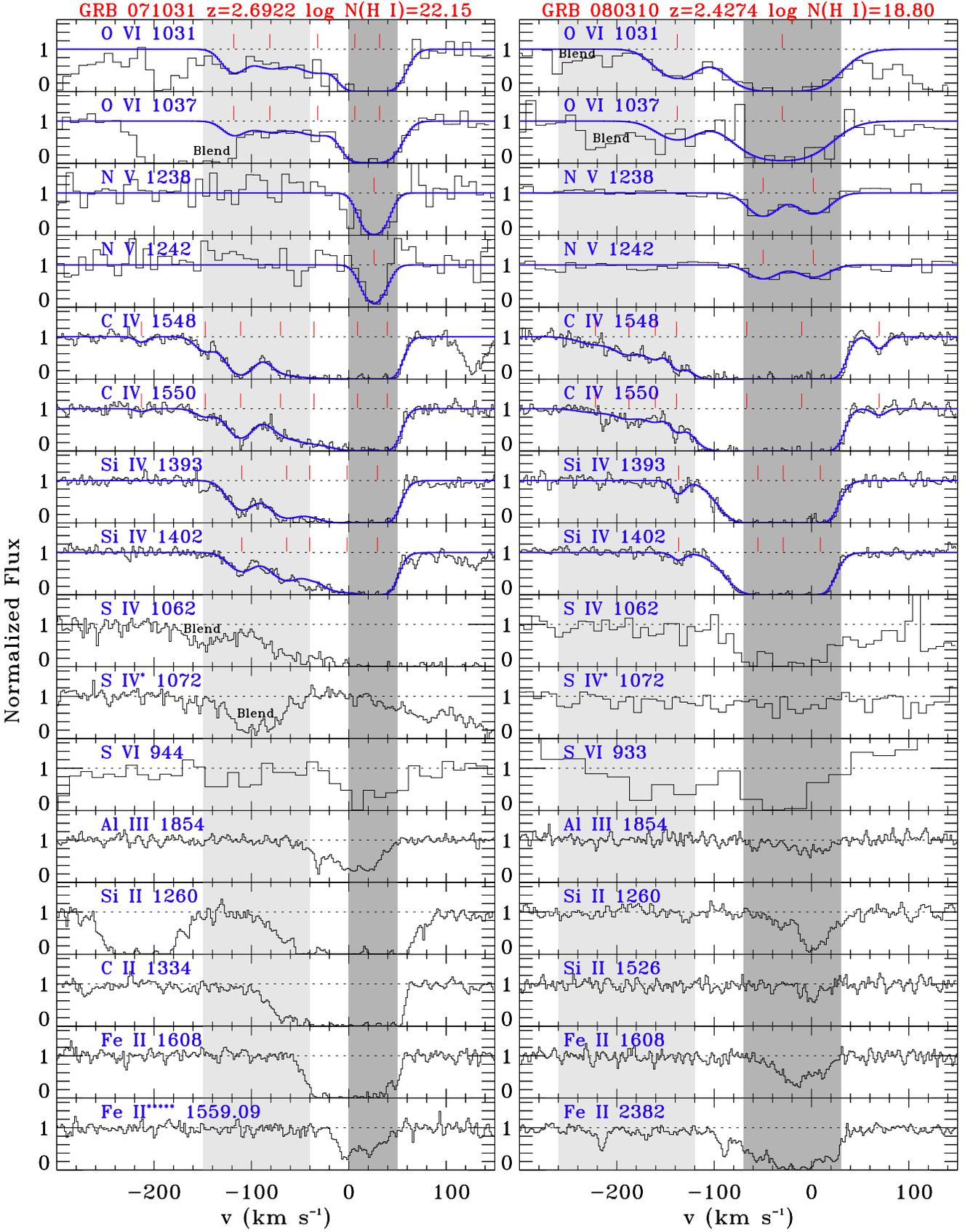}
\caption{Normalized high-ion and low-ion absorption-line profiles on a velocity
  scale relative to \zgrb\ for GRBs 071031 and 080310. Voigt profile
  fits are shown in blue (solid line), with the center of each
  component in the fit marked with a red dash. Dark shading denotes
  the regions of strong absorption, and light shading denotes the
  negative-velocity wings. The \os\ and \nf\ data have been rebinned
  by four pixels for GRB 071031 and by five pixels for GRB 080310. For
  \ssix, the rebin factors are five and ten, respectively. Absorption
  components at higher velocity relative to \zgrb\ are shown in Figure 5.} 
\end{figure*}

\begin{figure*}
\includegraphics[width=18cm]{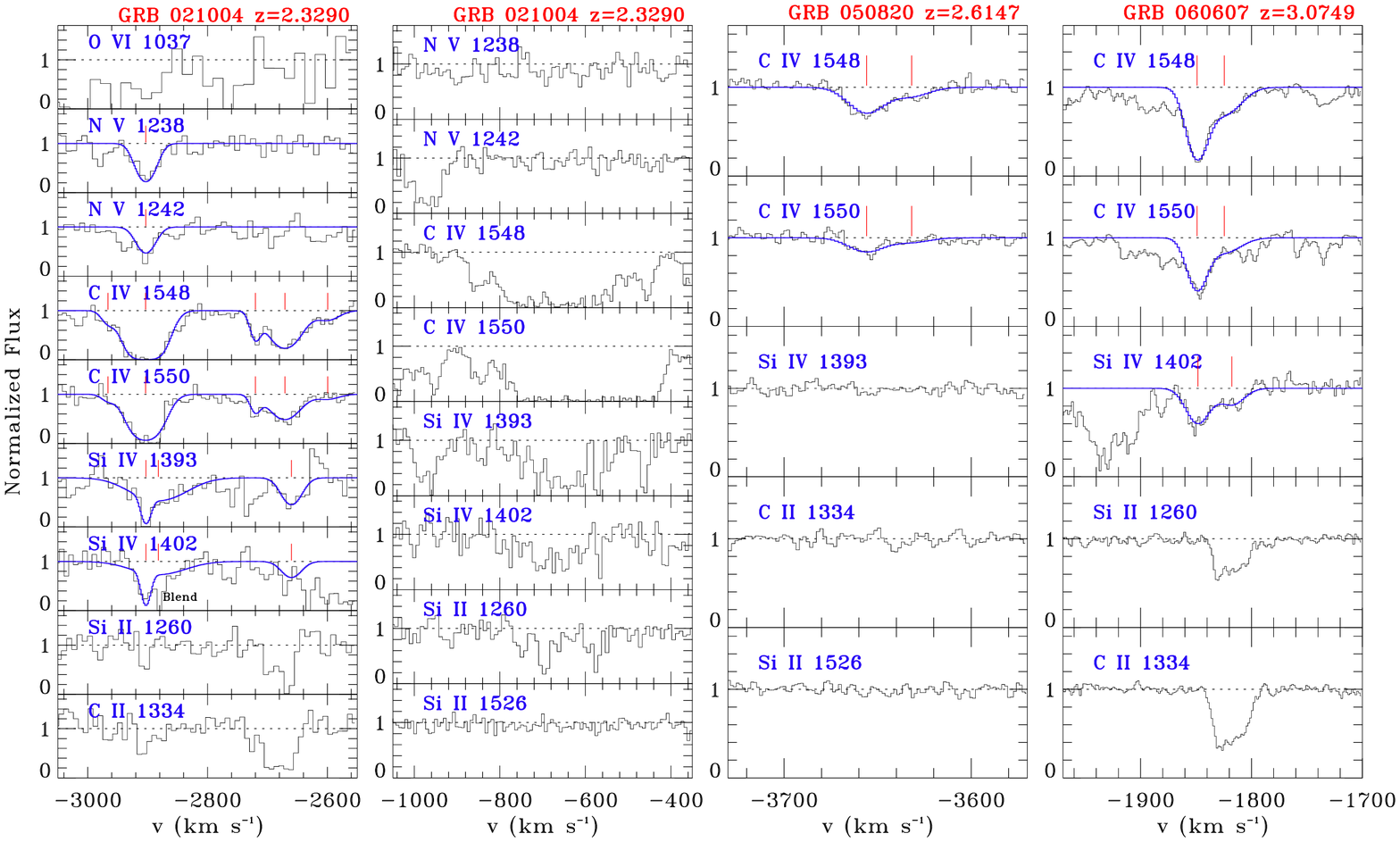}
\includegraphics[width=18cm]{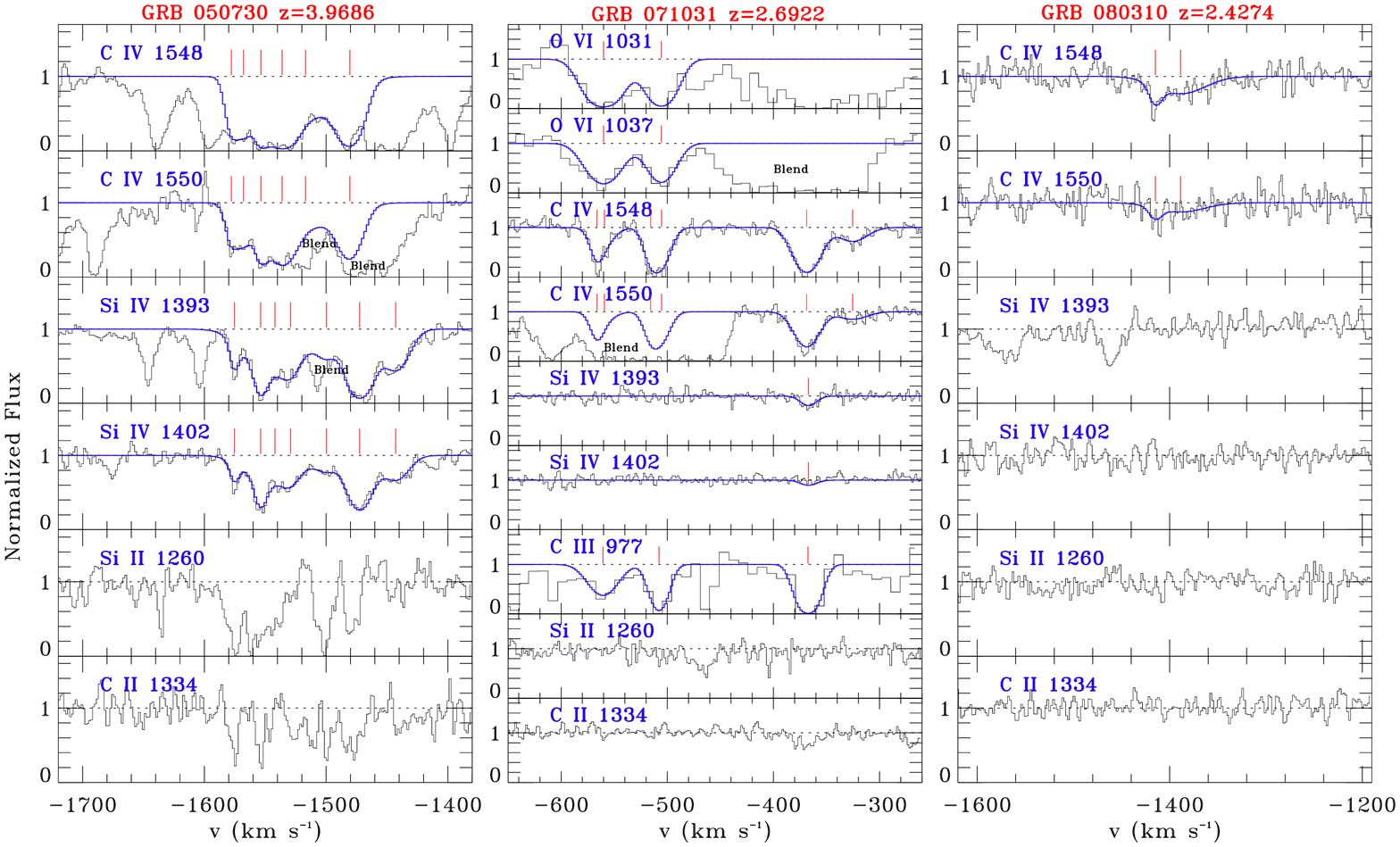}
\caption{Seven high-velocity (500--5\,000~\kms\ from \zgrb) 
  high-ion absorbers are seen in the seven GRB afterglow spectra in
  our sample; these HV absorbers are shown here
  (GRB~021004 shows two HV absorbers, GRB~050922C
  shows none, and the other five GRB spectra show one each).
  Voigt profile fits are only shown (with blue solid lines) for cases
  where we judge that reliable column densities can be extracted.
  The properties of these HV absorbers are diverse, and it is unlikely
  that they represent a single population. Several of these HV
  absorbers are seen in both neutral-phase and high-ion-phase
  absorption; others (e.g. the HV components
  toward GRB~071031 and GRB~080310) are seen only in high-ion transitions.}  
\end{figure*}

The detection rate of \nf\ in our sample is identical to that reported by
\citet{Pr08b}, who find \nf\ at \zgrb\ in 6/7 GRB afterglow spectra,
but since five systems are common to both samples, this is not surprising.
\nf\ has also been detected at \zgrb\ in the afterglow spectrum of
GRB~060206 \citep{Th08b}.
Other than the \os\ absorption in the GRB~050730 spectrum, which has been
published by \citet{DE07}, no reports of \os\ at \zgrb\ in GRB
afterglow spectra exist, and no previous reports of \sfo\ or \ssix\
exist either. Indeed, very few reports of interstellar \ssix\ absorption exist
anywhere in the high-$z$ \citep{CC99, Lv03} or low-$z$ \citep{Sa05b}
Universe. Interstellar \sfo\ absorption-line detections have
(to our knowledge) only been reported in quasar-intrinsic systems
\citep{Ar99, Ga06}. Finally, our non-detections of absorption in \sfo$^*$
$\lambda$1072.973 contrast with detections of this line
in the winds of OB-type stars \citep{Ma03, Le03}.

The absorption features observed at and near \zgrb\
can be classified into various types.
First, strong absorption is seen exactly at \zgrb\ in the form of
saturated \os, \cf, and \sif\ lines, usually with accompanying \nf,
and (where the data are unblended) also in \sfo\ and \ssix.
Second, complex multi-component \cf\ and \sif\ profiles with total velocity
widths of 100--400~\kms\ are seen in all seven GRB afterglow spectra.
Third, 4/7 GRB afterglow spectra show asymmetric negative-velocity
absorption-line wings extending over velocity regions of
65--140~\kms. These wings are seen clearest in \cf\ and \sif, but are
also present in \os. 
Fourth, 6/7 cases show HV components in \cf, and occasionally also in
neutral-phase lines such as \siw\ and \cw. 
Whereas the first three of these categories are visible 
in Figures 1 to 4, we present a separate plot (Figure 5) showing the
HV high-ion absorbers detected in our sample. We return to a
discussion of these categories of absorption in \S4,
after describing the details of the high-ion absorption in each spectrum. 
Unless otherwise stated, solar abundances were taken from \citet{Lo03}.

\subsection{GRB~021004 \zgrb=2.3290}
Extremely strong \cf\ absorption is seen at \zgrb, with a fully
saturated trough spreading over 300~\kms. 
\sif\ is also present and less saturated than \cf, but we cannot
discern the true component structure in \sif\ because of low S/N.
A component centered at 0~\kms\ is seen in \os, \nf, and
\sfo, with absorption also detected in \alth, \siw, \cw, and \few.
\sfo$^*$ $\lambda$1072.973 is not detected at \zgrb; we measure
log\,$N$(\sfo$^*$)$<$14.19 (3$\sigma$) over the velocity range $-$60 to 
60~\kms, and log\,$N$(\sfo)=15.30$\pm$0.29, 
so that $N$(\sfo$^*$)/$N$(\sfo)$<$0.08. 
Two HV absorbers are observed: one centered at $-$2\,900~\kms\ relative to
\zgrb, the other centered at $\approx\!-$600~\kms. The first is seen
in \sif, \cf, \nf, \siw, and \cw. The second is seen in \cf, \sif, and
\siw. These two HV absorption-line complexes are shown in separate
columns in Figure 5. The \os\ data are too blended in the HV
components to be of use. A full study of the UVES spectrum of
GRB~021004 is presented by \citet{Fi05}, and possible origins of the HV
components (e.g. a series of shells around the progenitor star)
are discussed in \citet{Mo02}, \citet{Mi03}, \citet{Sc03}, and \citet{La06}.
The velocity difference between the absorber at \zgrb\ and the HV absorber near
$-$500~\kms\ is equal to the velocity difference between the two lines
of the \cf\ doublet, i.e. line-locking is observed in \cf\ 
\citep[first noticed by][]{Mo02}\footnote{Line-locking is understood as
  occurring when a cloud is accelerated by radiation pressure in a
  given doublet line until the wavelength shifts into the shadow of
  the other doublet line cast by a separate cloud \citep{Vi99}.} .

\subsection{GRB~050730 \zgrb=3.9686}
High-ion absorption is seen in \os, \nf, \cf, \sif, \sfo, and \ssix\
centered at \zgrb. All of these lines except \nf\ are saturated.
The \nf\ profile shows two components separated by 27$\pm$5~\kms, each
with fairly narrow $b$-values (7$\pm$3 and 14$\pm$8~\kms); 
we fit the \ssix\ absorption with one component with $b$=17$\pm$2~\kms. 
\sfo$^*$ $\lambda$1072.973 is not detected at \zgrb; we measure
log\,$N$(\sfo$^*$)$<$13.70 (3$\sigma$) over the velocity range $-$25 to 
20~\kms, in which we measure log\,$N$(\sfo)=14.61$\pm$0.15, 
so that $N$(\sfo$^*$)/$N$(\sfo)$<$0.12. 
Note that this limit applies only to the strongest component of absorption.
The \cf, \sif, \os, and \cw\ profiles show a clear negative-velocity
absorption-line wing extending from $-$70 to $-$145~\kms.
Finally, a HV absorber at $-$1550~\kms\ is
seen with at least five \sif\ and \cf\ components spread over
$\approx$200~\kms\ \citep[see also][]{Ch07}.
This HV absorption-line complex is shown in Figure 5. 
This GRB shows an extremely high \hi\ column of log\,$N$(\hi)=22.15,
and is the highest redshift GRB in our sample.
A full study of the neutral and excited-state lines at \zgrb\
in the UVES spectrum of GRB~050730 is presented by \citet{DE07}.

\subsection{GRB~050820 \zgrb=2.6147}
\nf\ shows a single narrow component at $-$109~\kms\ 
relative to \zgrb\ with $b$=9$\pm$3~\kms.
This \nf-\zgrb\ offset is the largest among our seven spectra, 
and has been noticed before \citep{Pr08b}, but the formal significance 
of the $\lambda$1238.821 and $\lambda$1242.804 detections is
only 3.1$\sigma$ and 2.0$\sigma$ respectively. 
The \cf\ and \sif\ profiles exhibit at least ten components
spread over 400~\kms\ in velocity width, with a \cf/\sif\ ratio that
changes dramatically between components. \cf\ $\lambda$1550.781 is
blended so we only display \cf\ $\lambda$1548.204. A similar velocity
structure is observed at lower S/N in \alth\ $\lambda$1862.789.
\ssix\ $\lambda$933.378 shows three components aligned
with three components seen in \cf, strongly
suggesting the genuine presence of \ssix.
However, \ssix\ $\lambda$944.523 is affected by low S/N and possible
blending, so we treat the \ssix\ detection as marginal.
The \os\ profiles are highly blended and/or saturated, except for the 
range $-$140 and $-$200~\kms\ in \os\ $\lambda$1031.926, where the profile
closely follows the shape of the \cf\ and \sif\ profile. Therefore
while we are not able to measure the \os\ absorption in this case, we
know that it is present.
\sfo\ $\lambda$1062.662 shows a profile that broadly follows the \sif.
The \sfo$^*$ $\lambda$1072.973 profile is difficult to interpret. 
An absorption feature at 15~\kms does not correspond to any component 
in any other high ion, so we identify it as a blend. However, we
cannot rule out the presence of genuine \sfo$^*$ absorption between
$-$50 and 0~\kms.
Finally a two-component HV absorber at $-$3\,660~\kms\ relative to \zgrb\ is
seen in \cf, but with no detectable \sif, \cw, or \siw\ (see Figure 5).

\subsection{GRB~050922C \zgrb=2.1990} 
A single, strong component in \os, \nf, \cf\ and \sif\ is detected at
65~\kms\ relative to \zgrb. \alth\ is also detected in
this component in both $\lambda$1854.716 and $\lambda$1862.789. 
Our best-fit to the \os\ line has $b$=24$\pm$2~\kms; for the 
\nf\ line, we find $b$=16$\pm$3~\kms.
\cf\ and \sif\ show a complex series of at least
six components spread over $\approx$280~\kms, 
but with no evidence for a negative-velocity wing.
The \sfo\ and \ssix\ data are blended and not shown on Figure 3. 
This GRB shows a high log\,$N$(\hi) value of 21.60, and a low
metallicity of $-$2.0 \citep{Pr08b}. 
This is the only GRB afterglow spectrum in our sample in which no HV
\cf\ components are detected.

\subsection{GRB~060607 \zgrb=3.0749}
The high ions in the GRB~060607 spectrum behave differently than in the
other afterglow spectra. The \cf, \sif, and \sit\ profiles show a
four-component absorber centered at
\zgrb\ and covering only $\approx$100~\kms\ of total width. 
However, \nf\ is not detected at \zgrb; we derive a 3$\sigma$ upper limit
of log\,$N$(\nf)$<$12.70 in the velocity range $-$140 to 50~\kms,
where \cf\ absorption is seen. 
The \os, \sfo, and \ssix\ data are blended near \zgrb\ and are of no use.
\alth\ shows a non-detection.
The \cf\ profiles also show a weak (low optical depth) wing extending from
$-$60 to $-$120~\kms. 
This GRB has an extremely low \hi\ column density,
log\,$N$(\hi)=16.8 \citep{Pr08b}, 3.5 orders of magnitude lower than
the DLA columns seen in five of the other six cases.
A HV absorber seen in \cf\ and \sif\ is
centered at $-$1\,850~\kms\ relative to \zgrb, with nearby low-ion
absorption present in \siw\ and \cw (Figure 5).

We note that a sub-DLA at $z$=2.9372 is detected in the GRB~060607
spectrum with log\,$N$(\hi)=19.50$\pm$0.05 
\citep[first noticed by][]{Le06}. This sub-DLA is separated
from the nominal GRB redshift of 3.0749 by $\Delta z$=0.1377, or
10\,300~\kms, and shows strong, multi-component
absorption in \os, \cf, and \sif. We considered the
possibility that the host galaxy of GRB~060607 is actually located
at this redshift ($z$=2.9372), which would imply its 
ISM properties are more similar to those of the host galaxy ISM 
of the other bursts in our sample. However, this hypothesis is
unlikely to be correct since 
(a) \lya\ forest lines are detected up to the redshift of 3.0749, 
suggesting that \zgrb\ really is this high, and 
(b) this scenario would require the absorber observed at $z$=3.0749 to be
in front of the host galaxy yet moving toward it at over
10\,000~\kms, which is not easy to explain dynamically.
We conclude that the sub-DLA at 2.9372 traces a foreground galaxy
unrelated to the GRB. 

\subsection{GRB~071031 \zgrb=2.6922}
A strong component in \os, \nf, \cf, and \sif\ is detected at
\zgrb. The least-saturated high-ion line, \nf, shows 
a narrow line width of $b$(\nf)=11$\pm$2~\kms, implying
log\,$T\!<\!5.08$ for the gas containing the N$^{+4}$ ions, regardless
of ionization mechanism. 
No information is available on the line widths of \cf\ and \sif\ at
\zgrb\ because the absorption in these lines is saturated.
Whereas \ssix\ $\lambda$933.378 is blended, \ssix\ $\lambda$944.523
appears to show absorption at \zgrb, with a profile that matches
that of \os. However, because of the low S/N of the data at these
wavelengths, we do not attempt to derive a \ssix\ column density.
We measure log\,$N$(\sfo$^*$)$<$13.87 (3$\sigma$) over the velocity range 
0 to 50~\kms, but complete saturation (and possible blending) prevents
a reliable measurement of log\,$N$(\sfo) in this velocity range.
Absorption in several fine-structure lines of Fe, Ni, O, Si, and C is
detected at \zgrb\ in the GRB~071031 afterglow spectrum: in Figure 4
we show a line (\few$^{*****}$~$\lambda$1559.085) from the fifth
excited level ($^4$F$_{9/2}$) of \few. 
We derive an \hi\ column density log\,$N$(\hi)=22.15$\pm$0.05 and a
metallicity [Zn/H] of $-$1.72$\pm$0.06, where zinc is chosen since it
is typically undepleted onto interstellar dust grains.

A notable feature of the \cf, \sif, \sfo, and (at lower S/N)
the \os\ profiles is the wing extending blueward from 0~\kms\ down
to $-$160~\kms. A component at $-$115~\kms\ is superimposed on the wing. 
The \cf\ and \sif\ profiles follow one another closely, suggesting
these two ions form in the same physical regions of gas.
However, the slope d$N_{\rm a}(v)$/d$v$ is shallower for \os\ than for
\cf\ and \sif\ in the range $-$90 to $-$50~\kms\ (see \S4.3). 

Three strong components of HV high-ion absorption are seen
(Figure 5), at velocities of $-$560, $-$510, and $-$370 \kms.
The $-$560 and $-$510~\kms\ components are seen in 
\cf\ $\lambda$1548.204, \os\ $\lambda$1031.926,
\os\ $\lambda$1037.617, and \ct\ $\lambda$977.020, but are blended in \cf\
$\lambda$1550.781 and absent in \sif.
The $-$370~\kms\ component is detected in \ct, \cf, and \sif,
but the \os\ profiles are fully saturated at this velocity.
The \nf\ data show no significant detection in the HV gas, 
though the S/N near the \nf\ lines is too low to place a strong
constraint on $N$(\nf).
Neutral-phase absorption is not seen in the three HV
components, even in the sensitive (high $f$-value) lines
\ion{O}{i} $\lambda$1302.169, \siw\ $\lambda$1260.422, and \cw\
$\lambda$1334.532, with the single exception of a detection of weak
\cw\ at $-$370~\kms. The $N$(\cf)/$N$(\sif) ratios change dramatically
between the various components, from 4.8$\pm$1.0 at $-$110~\kms, to 
40$\pm$9 at $-$370~\kms, 
$>$140 (3$\sigma$) at $-$510~\kms\ and 
$>$110 (3$\sigma$) at $-$560~\kms. 

\subsubsection{Time-series observations of GRB~071031}

\begin{figure}
\includegraphics[width=9cm]{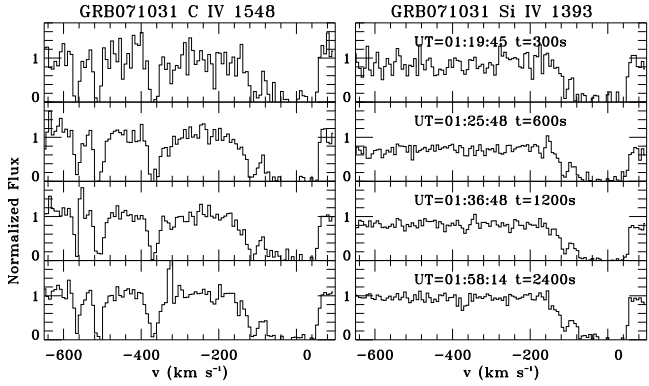}
\caption{Time-series observations of \cf\ and \sif\ absorption in the
  GRB~071031 afterglow spectra. Each row shows a different observing
  epoch, with the UT of the start of each
  exposure annotated with the exposure duration on the right panel. 
  These data have been binned by four pixels to improve the S/N.
  No evidence for time-variation in the high-ion profiles is seen.}
\end{figure}

The combined UVES spectrum of GRB~071031 (shown in Figure 4) 
was formed by co-adding four useful individual exposures, with
exposure times of 300, 600, 1200, and 2400~s (the sequence of
increasing exposure times was chosen to compensate for the
fading of the afterglow). Analysis of these
individual frames allows us to look for time-variation in the high-ion
absorption. We look for variation in the \cf\ and \sif\ profiles, since these
lines fall in parts of the spectrum where the S/N is highest
(a \nf\ time-series is unavailable). 
We focus on the HV components, since the high-ion absorption
at \zgrb\ is saturated, making time-variation at that redshift
difficult to detect. The time-series spectra are shown in Figure 6. 
No evidence for any evolution in the \cf\ or \sif\ profiles is
seen. The lack of variation can be quantified by measuring the
equivalent width at each observing epoch. The total equivalent width
and 1$\sigma$ error measurements in the range $-$600 to $-$330~\kms\ 
(over three HV components) are 266$\pm$88, 327$\pm$49, 233$\pm$36,
and 306$\pm$30~m\AA\ in the four exposures respectively, 
consistent with no variation in the absorption with time.
Thus the behavior of the high-ion lines in these HV absorbers is in
contrast to the time-variation in low-ion and fine-structure
absorption lines at \zgrb\ reported by \citet{DZ06} and \citet{Vr07}.

\subsection{GRB~080310 \zgrb=2.4274}
Strong high-ion absorption is seen in \os, \nf, \cf, \sfo, and \sif\
centered within 50~\kms\ of \zgrb.
The \nf\ profile shows two components separated by
50$\pm$3~\kms\ and with $b$-values of 18$\pm$3 and 19$\pm$3~\kms. 
The \ssix\ profiles are suggestive of absorption at \zgrb, but with
very low S/N and an inconsistency between the absorption in
$\lambda$933.378 and in $\lambda$944.523. For this reason, we do not
attempt to derive a \ssix\ column density. Similarly the \sfo\
$\lambda$1062.662 profile is too noisy to measure a reliable column, but
we can measure a lower limit of log\,$N$(\sfo)$>$14.77 over the
velocity range $-$70 to 30~\kms. Combining this with the constraint 
log\,$N$(\sfo$^*$)$<$14.38 (3$\sigma$) from the non-detection of
\sfo$^*$~$\lambda$1072.973, we find $N$(\sfo$^*$)/$N$(\sfo)$<$0.41. 
The \cf, \sif, and \os\ profiles show a clear negative-velocity
absorption-line wing extending from $-$120 to $-$260~\kms. 
This is the most extended absorption-line wing in our sample.
Finally, a HV absorber at $-$1400~\kms\ is seen (see Figure 5)
with two \cf\ components, one of which is very narrow
($b$=4$\pm$2~\kms). 
For this GRB-DLA we derive an \hi\ column density
log\,$N$(\hi)=18.80$\pm$0.10 and a metallicity 
[O/H] of $-$1.39$\pm$0.10. No ionization correction was applied since
a charge-exchange reaction closely links \ion{O}{i} and \hi, implying
[\ion{O}{i}/\hi]$\approx$[O/H] \citep{FS71}. 

\section{Discussion}
The six high-ion absorption lines reported in this paper 
at velocities near \zgrb\ are 
\sif, \sfo, \cf, \ssix, \nf, and \os, which trace the 
Si$^{+3}$, S$^{+3}$, C$^{+3}$, S$^{+5}$, N$^{+4}$, and O$^{+5}$
ions, requiring energies of 33.5, 34.8, 47.9, 72.6, 77.5, and 113.9~eV
for their creation, respectively \citep{Mo03}. 
A galactic ionizing spectrum, dominated by the integrated radiation
from O and B stars, drops strongly above 54~eV, 
the \ion{He}{ii} ionization edge \citep{BH86}. So whereas the ions
Si$^{+3}$, S$^{+3}$, and C$^{+3}$ can be photoionized in the ISM by starlight, 
the ions S$^{+5}$, N$^{+4}$, and O$^{+5}$ cannot.
We expect this to remain true in GRB host galaxies, even though 
they have been shown to harbor significant numbers of 
Wolf-Rayet stars \citep{Ha06}, since Wolf-Rayet spectra also show a
strong break at 54\,eV \citep{Cr07}. Instead, the detection of the
three ions S$^{+5}$, N$^{+4}$, and O$^{+5}$ implies the presence of
either a hard, non-stellar radiation source or hot, collisionally
ionized gas.

In the seven GRB afterglow spectra in our sample, the 
median and standard deviation of log\,$N$(\os) at \zgrb\ is 15.12$\pm$0.54.
For comparison, the median log\,$N$(\os) measured in 100 sight lines passing
through the Milky Way halo is 14.38 \citep{Wa03, Sa03}, rising to
14.80 when including the contribution from high-velocity clouds \citep{Se03}.
Turning to \nf, we report a median log\,$N$(\nf)=14.03$\pm$0.49 in the
seven GRB spectra, whereas the median log\,$N$(\nf) in 32 Galactic
halo sight lines is 13.45 \citep{IS04}.
High-ion absorption in \nf\ and \os\ has also been observed in 
$z$=2--3 DLA galaxies, with a median log\,$N$(\os)=14.77 \citep{Fo07a}, 
and 76 per cent of DLAs showing log\,$N$(\nf)$<$13.50 (Fox et
al. 2008, in preparation). 
Thus the column densities of \os\ and \nf\ seen at \zgrb\ are, in the
mean, higher than those seen in the halo of the Milky Way and those in DLAs.
The highly ionized Galactic ISM 
is a complex environment, with many physical
processes contributing to the production of the high ions, 
including conductive interfaces, turbulent mixing layers, shocks,
expanding supernova remnants, and galactic fountain flows 
\citep{Sa03, Zs03, IS04, Bo08}, and there is no reason why the 
ISM in the host galaxies of GRBs should be any less complex.

With this in mind, we now discuss the various categories of high-ion
absorption seen near \zgrb. Throughout this section, we refer the
reader to Table 9, which summarizes the high-ion to high-ion column
density ratios measured in the GRB afterglow spectra, and compares
them to the ratios measured in various other astrophysical environments.

\begin{table}
\begin{minipage}[t]{\columnwidth} 
\caption{High-ion column density ratios} 
\begin{tabular}{lccc}
\hline\hline
GRB\footnote{We report the observed column density ratios
  in different velocity regions: in the strong components seen at
  \zgrb, in the blueshifted wings, and in the HV
  absorbers. All limits are 3$\sigma$. In cases marked $^*$,
  saturation, blending, or low S/N prevents
  the ratio from being measured reliably.} &  
$\frac{N({\rm \cf})}{N({\rm \os)}}$ & 
$\frac{N({\rm \cf})}{N({\rm \sif)}}$ & 
$\frac{N({\rm \nf})}{N({\rm \os)}}$\\
\hline
\bf{At $z$(GRB)}\\
050922C   & $^*$         & $^*$         & 0.21$\pm$0.04\\ 
060607    & $^*$         & 7.6$\pm$1.2     &  $^*$     \\ 
\hline
\bf{Blueshifted Wings}\\
050730    & 0.40$\pm$0.25 & 3.7$\pm$0.9 & $<$0.09  \\ 
060607    & $^*$          & $>$13       & $^*$     \\ 
071031    & 1.6$\pm$0.4   & 3.2$\pm$0.9 & $<$0.11  \\ 
080310    & 0.55$\pm$0.11 & 25$\pm$4    & $<$0.05  \\ 
\hline
\bf{HV components}\\
021004 $-$2900~\kms\ & $^*$          & 5.1$\pm$3.1 & $^*$\\ 
050730 $-$1550~\kms\ & $^*$          & 3.7$\pm$0.4 & $^*$\\ 
050820 $-$3660~\kms\ & $^*$          & $>$24       & $^*$\\ 
060607 $-$1850~\kms\ & $^*$          & 3.2$\pm$0.5 & $^*$\\ 
071031 $-$560~\kms\  & 0.19$\pm$0.06 & $>$110   & $<$0.06\\ 
071031 $-$510~\kms\  & 0.35$\pm$0.10 & $>$140   & $<$0.07\\ 
071031 $-$370~\kms\  & $^*$          & 40$\pm$9    & $^*$\\ 
080310 $-$1400~\kms\ & $^*$          & $>$36       & $^*$\\ 
\hline
Galactic Halo\footnote{Mean$\pm$1$\sigma$ in 16 Milky Way halo sight lines
  \citep{Zs03}.}  
& 0.60$\pm$0.47 &  3.46$\pm$1.09 & 0.12$\pm$0.07\\
LMC\footnote{Range observed in LMC gas (not including \hw\
    regions) in four sight lines \citep{LH07}.} 
& $<$0.09--0.59 & 1.5--2.5 & $<$0.33\\ 
DLAs, $z$=2--3\footnote{Mean$\pm$1$\sigma$ in
10 DLAs \citep{Fo07a}.} 
& 2.1$\pm$0.7 & 5.4$\pm$2.2 & ... \\
IGM, $z$=2--4 & 0.35\footnote{Median value among four IGM absorbers
  measured by \citet{Si06}.}  
& 16$\pm$4\footnote{Median$\pm$1$\sigma$ 
in 188 IGM absorbers \citep{Bo03}.} & $<$0.23\footnote{Measured in
  composite QSO spectrum by \citet{LS93}.}\\
\hline
\end{tabular}
\end{minipage}
\end{table}

\subsection{Strong absorption at \zgrb}
We always detect strong high-ion absorption
exactly at \zgrb\footnote{In two cases the velocity centroid of the
  strongest high-ion absorption component differs from the nominal GRB redshift
  (measured from the low-ion lines) by more than 10~\kms: for
  GRB~050922C the offset is 65~\kms, and for GRB~071031 the offset is
  25~\kms. However, since the GRB redshifts are not known to better
  than a precision of several tens of \kms, these offsets are not
  significant, and we treat the strong high-ion absorption as arising
  exactly at \zgrb.}, 
in the form of saturated \os, \cf, and \sif\ components,
and (where the data are unblended) also in \sfo\ and \ssix.
\nf\ is present in 6/7 cases, with the advantage of being less
saturated than the other high-ion lines. Because of this, \nf\ is the
best line to investigate the line width and optical
depth in the strong absorption components. In the six cases where \nf\ is
detected, we fit eight \nf\ components, with $b$-values of 
7$\pm$3, 9$\pm$3, 11$\pm$2, 14$\pm$8, 16$\pm$3, 18$\pm$3,
19$\pm$3 and 28$\pm$3~\kms. The median of these eight values is 16~\kms. 
A \nf\ component formed in gas at 200\,000~K, the temperature at which
the production of \nf\ by collisions is maximized \citep{GS07}, 
has a $b$-value of 15.4~\kms. Therefore, we are unable to rule out
collisional ionization for the gas traced by \nf.

\subsubsection{Photoionized circumburst gas?}
Noticing that the \nf\ absorption at \zgrb\ tends to be stronger and narrower
(in total velocity width) than the \nf\ absorption observed in the
Milky Way ISM and in DLAs, \citet{Pr08b} have recently argued that the
\nf\ absorption components at \zgrb\ do not trace ISM material in the
halo of the host galaxy, but rather arise in circumburst gas in the
immediate vicinity of the GRB. Our high-resolution UVES dataset,
covering a wider range of ionization states than has been observed
before, allows us to investigate this intriguing idea.

Our data are qualitatively consistent with the circumburst
hypothesis in the cases of GRBs~021004, 050922C, and 071031, where we
see strong, single components of \nf\ at \zgrb, each with 
log\,$N$(\nf)$>$14.0, aligned with other high ions including \os\ with
log\,$N$(\os)$\ga$14.5. For these cases,
our new measurements of the \os, \sfo, and \ssix\ column densities at
\zgrb\ (listed in Tables 2, 4, and 7) could be used to constrain
photoionization models of the circumburst region. 
However, in the remaining four spectra in our sample
(GRBs~050730, 050820, 060607, and 080310) the situation is
more complex. In each of these four cases, log\,$N$(\nf)$<$14.1, and
in two cases log\,$N$(\nf)$<$13.2, i.e the observed \nf\ is \emph{not} in
excess of what is expected from the host galaxy's ISM.
In two of these cases (GRBs 050730 and 080310) the \nf\
absorption profiles show two components separated by a few tens of
\kms\ -- this is not the signature of a single burst. 
In another case (GRB~050820) there is an offset of 110~\kms\ between a
single \nf\ component and the strongest absorption in the other high ions.
And finally, in the case of GRB~060607, no \nf\ is detected at all.
In addition to these case-by-case details, we repeat that
the $b$-values of the detected \nf\ absorption components at \zgrb\
are \emph{not} particularly narrow, so photoionization is not required
by the line widths (though it is not ruled out either).

\subsubsection{Total ionized column density}
We can further investigate the nature of the strong components seen at
\zgrb\ by calculating the total column density of ionized hydrogen
$N$(hot~\hw)\footnote{Here we use the word ``hot'' simply to label the
  \nf-bearing gas; the upper limit on the temperature is
  215,000\,K for the median \nf\ $b$-value of 16~\kms.}
contained in these components. To do this we have to correct for ionization
and metallicity, so the calculation is only possible for cases where
the nitrogen abundance has been measured in the DLA at \zgrb. This is
true for three GRBs in our sample: [N/H]=$-$3.16$\pm$0.10 for GRB~050730,
$>\!-$1.35 for GRB~050820, and $<\!-$4.09 for GRB~050922C \citep{Pr07b}. 
The relationship between the \nf\ and hot \hw\ column densities can be
written in the following way:
\begin{equation}
N({\rm hot~\hw})=\frac{N{\rm (\nf)}}{{\rm (\nf/N)(N/H)_n}}\frac{{\rm
    (N/H)_{n}}}{({\rm N/H)_{\rm i}}}, 
\end{equation}
where (N/H)$_{\rm i}$ denotes the nitrogen abundance in the ionized gas
and (N/H)$_{\rm n}$ denotes the nitrogen abundance in the neutral gas
(which is measured).
We take the solar nitrogen abundance of (N/H)$_\odot=10^{-4.22}$
from \citet{Gr07}, and then (N/H)$_{\rm n}=10^{\rm [N/H]}$(N/H)$_\odot$. 
If the gas is collisionally ionized, as would be
appropriate for \nf\ in the host galaxy ISM, \nf/N$<$0.25 at all
temperatures \citep{GS07}\footnote{This is true for either equilibrium
  or non-equilibrium models. However, in the \citet{Pr08b}
  GRB photoionization models, a \nf\ ionization 
  fraction as high as 0.6 is predicted.}. 
Therefore we adopt 0.25 as the maximum allowed ionization
fraction, corresponding to a minimum ionization correction of a factor of 4.
If the ionized gas has the same metallicity as the
neutral gas, so that the term (N/H)$_{\rm n}$/(N/H)$_{\rm i}$ in
Equation 1 is equal to one, then $N$(hot~\hw) is:\\
$>$21.8 for GRB~050730 [which has log\,$N$(\hi)=22.15],\\
$>$19.3 for GRB~050820 [which has log\,$N$(\hi)=21.00], and \\
$>$22.7 for GRB~050922C [which has log\,$N$(\hi)=21.55],\\
where the results are lower limits since (a) we have used the smallest 
allowed ionization correction, and (b) they only apply to the
\nf-bearing gas, and \hw\ may exist at other temperatures as well.
Comparing the values of $N$(hot~\hw) and $N$(\hi), we find 
the hot-ionized-to-neutral ratio $N$(hot~\hw)/$N$(\hi) takes values of
$>$0.4 and $>$0.02 for GRBs~050730 and 050820, and $>$13 for
GRB~050922C. Whereas the first two values are modest, the
GRB~050922C value is extremely (and implausibly) high -- it
requires over ten times as much mass in hot ionized gas in a single
component as there is in neutral gas in the entire host galaxy. 
However, if the nitrogen abundance in the ionized gas was higher than in
the neutral gas, so that (N/H)$_{\rm n}$/(N/H)$_{\rm i}<1$, then solutions
with lower $N$(hot~\hw) would be possible. Thus our calculation of the
total ionized column density at \zgrb\ suggests that the
metallicity in the highly ionized gas is higher than the metallicity in the
neutral gas (particularly for GRB~050922C); however, this calculation
is inconclusive in determining whether the gas is circumburst or interstellar.

\subsubsection{Photo-excitation modeling of $N$(\sfo) and $N$(\sfo$^*$)}
\begin{figure}
\includegraphics[width=9cm]{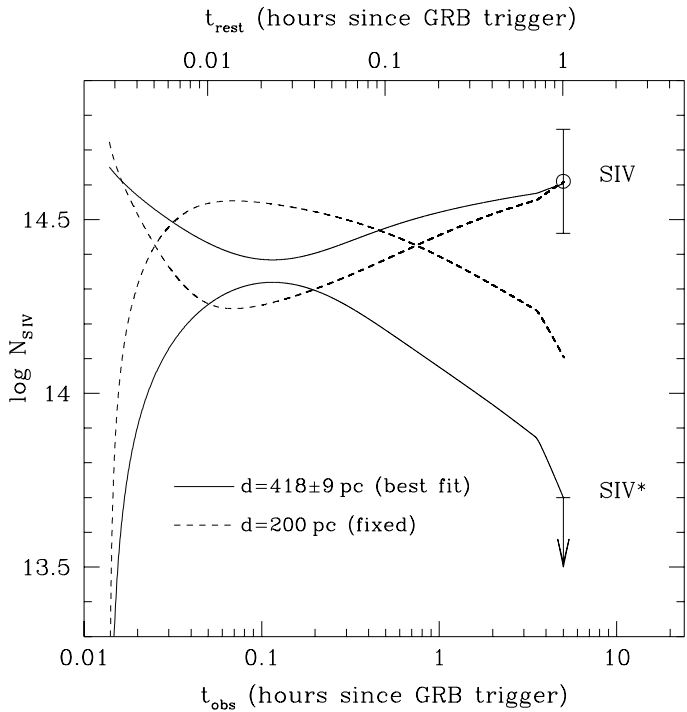}
\caption{Analysis of the \sfo\ level populations at \zgrb\ toward
  GRB~050730. The solid and dashed lines show the predictions from
  UV photo-excitation models for the time-variation of the \sfo\ 
  and \sfo$^*$ column densities (see text for details). 
  The data points show the observed $N$(\sfo) 
  and upper limit to $N$(\sfo$^*$) in the UVES spectrum, plotted at
  the $x$-position corresponding to the mid-observation epoch 
  (five hours after the trigger).
  The solid lines show the model with the best-fit distance of 418$\pm$9~pc. 
  The dashed lines show a model with the distance $d$ fixed at 200~pc,
  for illustration. In this case $N$(\sfo$^*$) would be 0.5 dex stronger than
  observed, so this model is clearly ruled out. The result that the
  \sfo-absorbing gas must lie at $d>400$~pc implies an interstellar
  rather than circumburst origin.}
\end{figure}

Another method for distinguishing between circumburst gas and more
distant interstellar gas is to look for absorption from 
fine-structure levels. In circumburst gas, UV pumping by GRB photons will
populate the excited electronic levels, which decay and cascade to
populate the fine-structure levels of the ground state
\citep{Pr06, Ch07, Vr07}. Because of the rapid fading of the
afterglow, the level populations evolve in a non-equilibrium,
time-dependent manner. Our data allow us to investigate this
process in the \sfo\ ion, since in four GRB spectra 
the combination of a detection of \sfo\ $\lambda$1062.662 and a
non-detection of \sfo$^*$~$\lambda$1072.973 allows us to 
place an upper limit on the \sfo$^*$/\sfo\ ratio at \zgrb.
The Einstein $A$-value of the forbidden transition between 
the ground state of \sfo\ and its fine-structure
level is $7.7\times10^{-3}$~s$^{-1}$  
\citep[taken from NIST;][]{Ra08}\footnote{See
 http://physics.nist.gov/PhysRefData/ASD/index.html}.
In other words, in the absence of stimulated photo- and collisional
de-excitation, the fine-structure level probed by the \sfo$^*$
1072.973 transition will decay in 1/$A$=130\,s.
In fully ionized gas (as expected near GRBs) there is no \lya\ opacity
and so the \sfo\ lines (which lie shortward of \lya) will
not be shielded from UV radiation.
For GRBs~021004, 050730, and 080310 we measure 3$\sigma$
limits to the $N$(\sfo$^*$)/$N$(\sfo) ratio of $<$0.08, $<$0.12,
and $<$0.41 respectively (for the other four afterglows we are unable to
constrain this ratio due to blending). 

Following the technique described in \citet{Vr07}, we can model
the \sfo\ excitation level in gas subject to UV radiation from the GRB,
assuming that photo-excitation is the dominant excitation process. 
We chose to model the \sfo\ level population measured toward
GRB~050730, because this case has the strongest constraint
on $N$(\sfo$^*$). The model has five free parameters: the distance
$d$ between the GRB and the absorbing gas (assuming all the 
\sfo\ arises at the same distance), the total \sfo\ column 
density $N_{\rm tot}$ (equivalent to the pre-burst ground-state \sfo\
column density), the afterglow spectral slope $\beta$ (where
$F_\nu\propto\nu^{\beta}$), the Doppler $b$-parameter of the
absorbing gas, and the rest-frame time when the calculation begins
$t_0$ (i.e. the time when the afterglow photons begin to be tracked in
the code). For a given set of these parameters, the model predicts the
\sfo\ and \sfo$^*$ column densities as a function of time
(both in the observer frame and in the rest frame).
Since the time of mid-observation of GRB~050730 is known, five hours
(observer frame) after the trigger, predicted values for $N$(\sfo) and
$N$(\sfo$^*$) can be extracted from each model. 

The parameters in our GRB~050730 model were determined as follows.
$N_{\rm tot}$ was chosen to equal $N$(\sfo)+$N$(\sfo$^*$), assuming the
actual $N$(\sfo$^*$) is equal to the measured upper limit 
(if the actual value is lower, our conclusions are strengthened;
see below). The value for $\beta$ ($-$0.56) was determined using the
observed light curve for this afterglow.
The parameter $t_0$ is poorly constrained, but our runs showed
that the \sfo\ level populations are very insensitive to $t_0$ for
values between 0 and 300~s; we used a value $t_0$=10\,s \citep[see][]{Vr07}.  
Finally a $b$-value of 10~\kms\ was chosen as a typical value for
interstellar gas, and indeed matches (within the 1-$\sigma$ error) the
measured $b$-value for the \sfo\ component at 0~\kms\ toward
GRB~050730. This leaves the distance $d$ as the one genuinely floating
parameter in the model; a grid of models at different distance $d$ was
generated, and the predicted and observed $N$(\sfo) and
$N$(\sfo$^*$) were compared using a chi-squared minimization routine.

The model results are shown in Figure 7. 
The best-fit distance $d$ is 418$\pm$9~pc. 
Because we assumed that the actual \sfo$^*$ column density
was equal to the measured upper limit, this represents a strong
\emph{lower limit} of 400~pc (2$\sigma$) on the distance of the
absorbing gas from the GRB. In other words, if the gas was at
$d\!<\!400$~pc, we would have seen \sfo$^*$ absorption, whereas we
clearly do not. This represents an important result, since it implies
that the \sfo\ absorption has an interstellar rather than circumburst origin.
The distance limit on the \sfo\ absorption does not necessarily have
to apply to the \nf\ absorption (or the other high ions). 
However, the \nf\ and \sfo\ profiles
at \zgrb\ toward GRB~050730 are similar, both showing the
component near 0~\kms\ and a weaker component at 25~\kms, suggesting
that the two ions are co-spatial and hence that the \nf\ absorption
also arises at $>$400~pc from the GRB.

\subsubsection{Origin of strong absorption}
In summary, we find that while the circumburst hypothesis
explains the observation that in three afterglow spectra, strong
single-component \nf\ is aligned with saturated absorption in the
other high ions, it is challenged by the \nf\ profiles in the other four
datasets, it is not strictly required by the \nf\ line widths, and the
expected high-ion fine-structure lines and time-variable column
densities are yet to be observed. Furthermore, in one case (GRB~050730)
we are able to place a lower limit on the distance to the high-ion
absorbing gas of 400~pc, using the simultaneous detection of \sfo\
and non-detection of \sfo$^*$.
Therefore, based on our current dataset, we cannot rule out an 
interstellar explanation for the strong high-ion components at \zgrb,
noting that since the higher $N$(\hi) values in GRB-DLAs than in
QSO-DLAs suggest that GRB sightlines pass through the inner star-forming
regions of the host galaxies \citep{Pr08a, Fy08}, one would
\emph{expect} high column densities of interstellar plasma at
\zgrb, so the strength of the \nf\ absorption does not, in our view,
argue against an interstellar origin. 
We also note that a strong, apparently narrow (though unresolved) \nf\
component is observed in the spectrum of LBG cB58 at the galaxy redshift 
\citep[$z$=2.73;][]{Pe02}, showing that such components can be formed
even in the absence of a GRB.
Time-series observations of \nf\ in GRB afterglow spectra, which
unfortunately are not available in the current dataset, would help to
resolve this issue. A significant detection of time-variation in the
\nf\ column density at \zgrb\ \citep[see models by][]{Pr08b}
would represent the discovery of gas that is unambiguously close to
the GRB. 

\subsection{Multi-component absorption in \cf\ and \sif}
The multi-component structure seen in \cf\ and \sif\ spreading over
several hundred~\kms\ around \zgrb\ is not seen in the other high
ions. However, the \cf\ and \sif\ profiles lie out of the \lya\ forest, so 
are free from blending and show higher S/N than \os\ or \nf, 
and it is unclear how the \os\ and \nf\ profiles would appear if
observed at the same S/N. These complex, multi-component \sif\ and \cf\
profiles are reminiscent of the profiles
of these ions in damped \lya\ (DLA) absorbers
\citep{Lu96, Le98, WP00, Fo07b, Le08}, which
represent galaxy halos seen in absorption toward a background source. 
The median (mean) value of the total \cf\ velocity width
in 74 DLAs and sub-DLAs at $z$=2--3 reported by \citet{Fo07b} is
255~\kms\ (342~\kms), where the width is the total observed range of
absorption, regardless of optical depth. In the seven GRB spectra in
our current sample, the median (mean) total \cf\ velocity width
measured in an identical manner is 280~\kms\ (320~\kms), 
similar to the DLA value
(note we have excluded the HV components, which we treat separately).
A precise comparison of the median \cf\ column density between
the GRB sample and the DLA sample is not possible, because these
measurements are affected by saturation, but we can state that
log\,$N$(\cf) at \zgrb\ is $\ga$15 in all seven GRB spectra, which
is significantly stronger than the median DLA value
log\,$N$(DLA \cf)=14.27$\pm$0.57 measured by \citet{Fo07b}.
Nonetheless, the similarities in total \cf\ line width between GRB and
DLAs support a galactic origin for the multiple components seen in
\cf\ and \sif\ in a velocity range of several hundred \kms\ around \zgrb.

\subsection{Negative-velocity (blueshifted) absorption-line wings}
One notable feature present in four of the seven GRB afterglow spectra
in our sample is 
a negative-velocity absorption-line wing seen in \cf, \sif, and 
(at lower S/N) \os. These wings are characterized by relatively smooth,
asymmetric absorption extending for 65--140~\kms\ 
to negative velocities relative to the strong absorption at \zgrb.
To explore the high-ion velocity structure in the wings, 
we present in Figure 8 apparent column density profiles of the high
ions over the velocity of the wings, calculated using
$N_a(v)=3.768\times10^{14}\tau_a(v)/f\lambda$, where 
the apparent optical depth
$\tau_a(v)={\rm ln}\,[F_c(v)/F(v)]$, $F(v)$ and $F_c(v)$ are the
observed flux level and estimated continuum level
as a function of velocity, and where $\lambda$ is in \AA\
\citep{SS91}. We follow the error treatment of \citet{SS92}.
This form of display allows the velocity structure seen in each ion to
be closely compared. In two cases we add the profile of \cw\
$\lambda$1334.532, which also shows wing absorption. The high-ion column
densities in the wings are summarized in Table 10.

\begin{table*}
\begin{minipage}[t]{18cm}
\caption{Properties of absorption in the blueshifted wings}
\begin{tabular}{lcccc ccc}
\hline\hline
GRB & \zgrb\ & $v_{\rm min}, v_{\rm max}$\footnote{On velocity scale relative to \zgrb.}
 & $\delta v$\footnote{Velocity extent of wing $\delta v\!\equiv\!v_{\rm max}$--$v_{\rm min}$, shaded light grey on Figures 2, 3, and 4.}
 & log\,$N$(\os) & log\,$N$(\cf) & log\,$N$(\sif) & log\,$N$(\nf)\\
    & &       (\kms) & (\kms) & ($N$ in \sqcm) &
($N$ in \sqcm) & ($N$ in \sqcm) & ($N$ in \sqcm)\\
\hline
 050730 &  3.9686 & $-$145,$-$ 80 &  65 &                 14.47$\pm$0.35 & 13.79$\pm$0.03  &                13.21$\pm$0.02 &             $<$13.56 \\
 060607 &  3.0749 & $-$140,$-$ 60 &  80 &                            ... & 13.00$\pm$0.02  &                      $<$12.27 &             $<$12.91 \\
 071031 &  2.6922 & $-$150,$-$ 40 & 110 &                 14.25$\pm$0.06 & 14.45$\pm$0.09  &                13.95$\pm$0.06 &             $<$13.59 \\
 080310 &  2.4274 & $-$260,$-$120 & 140 &                 14.31$\pm$0.08 & 14.05$\pm$0.02  &                12.65$\pm$0.06 &             $<$13.27 \\
\hline
\end{tabular}
\end{minipage}
\end{table*}

\begin{figure*}
\includegraphics[width=18cm]{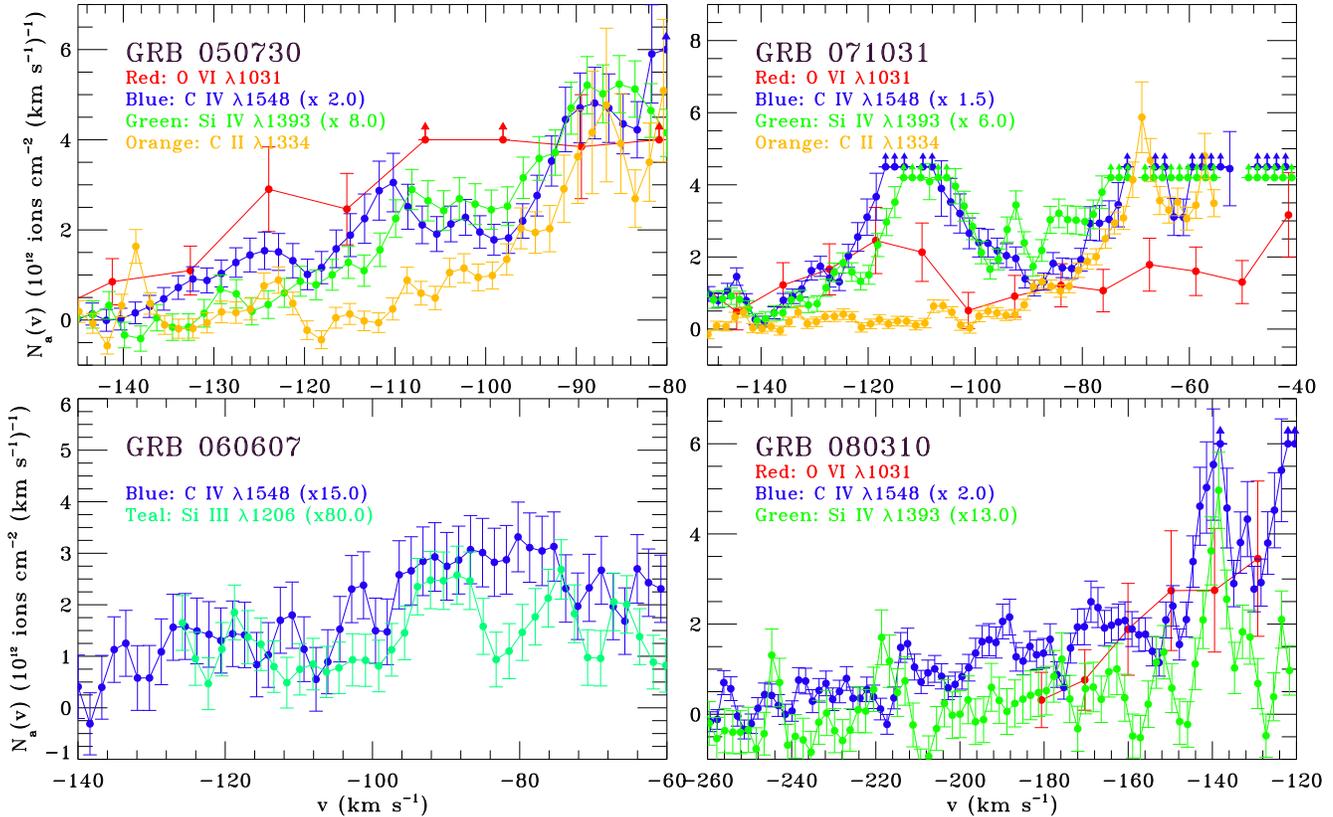}
\caption{High-ion apparent column density profiles
  in the four negative-velocity wings in our sample.
  Errors reflect both statistical noise and continuum placement uncertainties.
  The \cf\ and \sif\ profiles have each been scaled by the factors
  annotated on the plot to allow for comparison of their shape with \os.
  Only lower limits (arrows) can be presented
  for saturated pixels. The \os\ data have been rebinned by four
  pixels in the case of GRB~071031, eight pixels in the case of 
  GRB~050922C, and five pixels in the other cases.
  \nf, \sfo, and \ssix\ are not detected in the wings.}  
\end{figure*}

We find that the \cf\ and \sif\ profiles track one another closely in 
three of the four wings, indicating that the two ions form in the same
regions of gas. The exception is the very weak wing seen toward
GRB~060607, in which \sif\ is not detected, but \sit\ is seen.
For the wings seen toward GRBs~050730 and 080310, the \os\ and \cf\
profiles share a common gradient d$N_a$/d$v$. However, for GRB~071031,
the \os\ absorption shows a shallower slope. 
We measure a $N$(\cf)/$N$(\sif) ratio of $\approx$3 in two of the four
wings (GRBs~050730 and 071031), similar to the Galactic halo average of 
3.46$\pm$1.09 measured by \citet{Zs03}.
This supports an interstellar origin for the absorption-line wings.
In the wings toward GRBs~060607 and 080310 the $N$(\cf)/$N$(\sif) ratio is much
higher, taking values of $>$13 (3$\sigma$) and 25$\pm$4 respectively. 

\os\ absorption-line wings with similar velocity extent have been
observed at zero redshift in the ultraviolet spectra of quasars taken
with the \emph{Far-Ultraviolet Spectroscopic Explorer (FUSE)} satellite
\citep{Se01, Se03, Fo05, Fo06, Sa05a, Ke06}\footnote{Analogous \cf\
  and \sif\ wings are yet to be observed at zero redshift, but these
  two ions fall in the near-UV, requiring \emph{HST} observations
  rather than \emph{FUSE}, and the number of high-resolution QSO
  spectra taken with \emph{HST} is small compared to the number of 
  extragalactic sight-lines observed with \emph{FUSE}.}. 
\citet{Fo06} measured eleven Galactic \os\ wings,
and report a mean and standard deviation \os\ column density
of log\,$N$(\os)=13.85$\pm$0.45, with each wing extending over
$\sim$100~\kms; these authors successfully modeled the \os\
absorption-line wings as tracing Galactic outflows moving under
ballistic trajectories.
Furthermore, a strong \os\ wing covering a similar velocity range  
is seen in a sub-DLA (an absorber with
19.0$<$log\,$N$(\hi)$<$20.3) at $z$=2.67 \citep{Fo07c}.
The \cf\ profile observed in the LBG cB58 at $z$=2.7 shows a clear
absorption wing extending from $-$250 to $-$750~\kms\ 
($\approx$ nine resolution elements)
in the galaxy rest-frame \citep{Pe02}, and an asymmetric blueshifted
absorption feature in \ion{Mg}{ii} has been reported in the composite
spectrum of 1400 $z\!\approx\!1.4$ galaxies by \citet{We08}, who also
interpret it as a galactic outflow signature.

\begin{figure*}
\includegraphics[width=18.5cm]{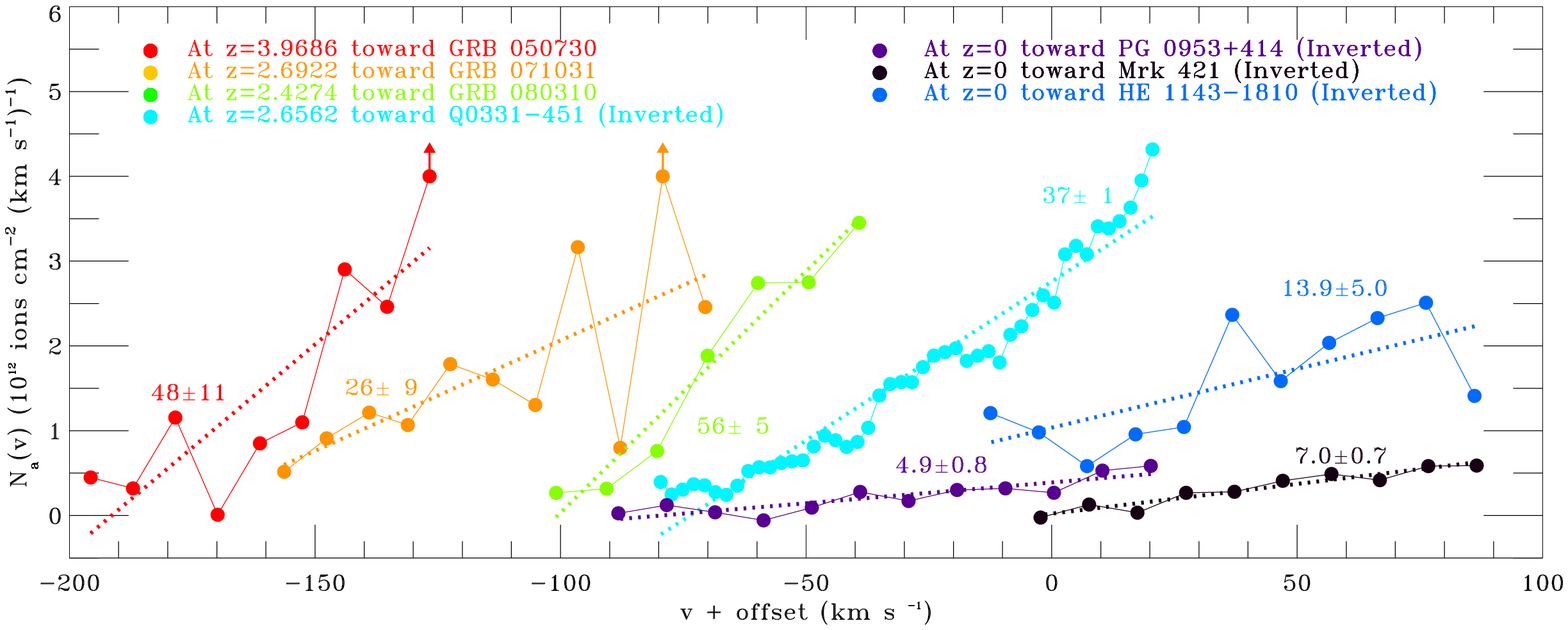}
\caption{Montage of apparent column density profiles of \os\
  absorption-line wings measured in various environments: near \zgrb\
  in three GRB afterglow spectra (this paper, three left-most lines), 
  at zero redshift in the Milky Way halo in the UV spectra of three
  AGN \citep[][ three right-most lines]{Fo06}, and in a
  sub-DLA at $z$=2.6562 \citep{Fo07c}. The Milky Way and sub-DLA cases
  have been inverted and velocity offsets have 
  been added for ease of comparison, but the gradient of each profile
  is preserved. Linear fits are shown with color-coded dashed lines,
  and the gradient of each fit in units of $10^9$~ions~\sqcm~(\kms)$^{-2}$
  is annotated with its 1$\sigma$ error on the panel.
  In these diverse galactic environments sampling over twelve billion
  years, the \os\ absorption-line profiles can be used to trace and
  quantify galactic outflows.} 
\end{figure*}

Whereas the $z$=0 \os\ wings are always detected at \emph{positive}
velocities (redshifted) relative to the Local Standard of Rest (LSR), 
the GRB wings are seen at \emph{negative} velocities (blueshifted)
relative to \zgrb. 
This behavior is strongly suggestive of galactic outflows:
in the Milky Way case (at least for high-latitude and anti-center sight
lines, where Galactic rotation effects are insignificant),
outflowing gas appears redshifted relative to the LSR, whereas in the
GRB case, outflowing gas moves toward us along the line-of-sight and
hence appears blueshifted relative to the host galaxy.
The possibility remains that some wing absorbers represent chance
alignments of unrelated components separated by small velocities,
which are unresolved by the spectrograph.
However, we find this explanation unlikely for the strongest wings
in our sample, seen toward GRBs~071031 and 080310, since in these
cases the column density in the wing increases 
monotonically with velocity\footnote{In the case of GRB~071031, the
  \cf\ profile shows a smooth gradient with velocity once the
  $-$115~\kms\ component has been subtracted off.}
in an interval of over 100~\kms, and there is no reason why a series
of random components should be aligned in this way.

To illustrate the similarity in the wing absorption features observed at
both low and high redshift, we show in Figure 9 a comparison of the
\os\ apparent column density profiles in the zero redshift and GRB
sight line wings. We have inverted the positive-velocity wings to
compare their gradient with the wings observed in the GRB spectra. 
While the similarity in the overall
shapes of the wing profiles is striking, the high-$z$ wings in the GRB
host galaxies show higher \os\ column densities (by a factor of $\approx$5) 
and steeper gradients d$N_{\rm a}$/d$v$ (by a factor of 5--10) 
than the Milky Way wings, implying higher mass flow rates.

\subsection{High-velocity components at 500--5\,000~\kms}
In six of our seven GRB spectra (all GRBs except 050922C), high-ion
components at 500--5\,000~\kms\ relative to \zgrb\ are observed (see
Figure 5). Such HV components have been observed before
in GRB~020813 \citep{Ba03},
GRB~021004 \citep{Mo02, Mi03, Sc03, Fi05, La06}, 
GRB~030226 \citep{Kl04}, 
GRB~050505 \citep{Be06}, and 
GRB~050730 \citep{DE07, Ch07}.
\citet{Ch07} report that \cf\ components with $W_{\rm r}\!>\!200$~m\AA\  
occur at 1\,000--5\,000~\kms\ in 20 per cent of GRB afterglow spectra.
In our data, the incidence of HV \cf\ absorbers (6/7, or 86\%) 
is much higher because (a) we have a higher sensitivity: the weakest
HV \cf\ absorber in our sample shows $W_{\rm r}$=48~m\AA, 
and (b) we classify absorbers in the range 500--1\,000~\kms\ as HV.

The HV absorbers seen in GRB afterglow spectra
do not represent a homogeneous population, as do the blueshifted
absorption-line wings and (potentially) the strong, narrow absorbers at \zgrb.
On the contrary, we observe a broad range of properties among the HV
absorbers. In three sight lines (toward GRBs~021004, 050730, and 060607), 
the HV components exhibit low-ionization absorption 
(clearest in \cw\ $\lambda$1334.532 and \siw\ $\lambda$1260.422).
However, the HV components toward GRBs~050820, 071031 and 080310 show no
(or very weak) neutral-phase absorption, in either \cw\ or \siw. So whereas the
presence of neutral gas in the first three cases implies the absorbing
material is not formed in the circumburst medium 
\citep[because this medium is expected to be fully ionized by the
  burst; e.g.][]{Vr07, Ch07, Pr08b}, we find that the properties of the
HV components in the latter three cases \emph{are}
consistent with an origin in the close circumburst medium.
This is particularly true for the HV gas toward GRB~071031, which
is detected in the form of strong, \os\ and \cf\
components that are not only absent in \cw\ and \siw, but also 
show non-detections of \sif. We have placed limits of
$N$(\cf)/$N$(\sif) $>$110 at $-$560~\kms, and $>$140 at $-$510~\kms\
in the GRB~071031 spectrum. These extreme ratios are far higher than
typical interstellar values \citep[Galactic ISM values for this ratio
  are $\approx$3;][]{Zs03} implying these HV components have a
different, non-interstellar origin. 

Given the connection between GRBs and massive stars \citep{MW99, WB06},
a leading explanation for the HV components in GRB afterglow spectra
is that they trace outburst episodes from the massive-star
progenitors of the GRBs \citep[e.g.][]{Sc03}, and thus arise in the
circumburst region. As pointed out by \citet{Be06}, a Wolf-Rayet
outflow model would give rise to absorbers that are enriched in carbon
and deficient in silicon, naturally explaining the
high \cf/\sif\ ratios measured in the HV absorbers.

If the wind-blown bubbles around GRB progenitors
are similar in size to those observed around Wolf-Rayet stars in the
Milky Way, their typical radius will be $\approx$2--10~pc \citep{Gr00}. 
Several groups have recently modeled Wolf-Rayet winds and their
relationship to the HV components in GRB afterglow spectra
\citep{Ra01, Ra05, Cv04, El06}, and in particular, predictions for the
high-ion signature of Wolf-Rayet winds are given by \citet{vM05,vM07,vM08}. 
Depending on observing epoch and viewing angle, 
components at 150-700~\kms\ and 1\,000-2\,000~\kms\ are
predicted at various stages of Wolf-Rayet evolution \citep{vM05};
these velocities match the observed range of HV components in our sample.
Recent models of the absorption-line signatures of galactic winds 
\citep[as opposed to single-star winds;][]{Fa07, KR07, Sa08} generally
predict high-ion components at velocities of only 100--200~\kms\
relative to the galaxy, though observations support the idea that
galactic winds can reach speeds of $\approx$1\,000~\kms\ \citep{Fr02,
  Tr07, We08}. Nonetheless, the fairly narrow $b$-values measured in
the HV components toward GRBs~071031 and 080310 
(not to mention the lack of \sif) are at odds with the
broader, collisionally ionized components expected in galactic winds
\citep{OD06}. Thus we conclude that among the available models,
Wolf-Rayet winds are the best explanation for the highly ionized HV
components toward GRBs~071031 and 080310.

\section{Summary}
We have performed a systematic study of the high-ion absorption near
\zgrb\ in seven GRB afterglow spectra observed with UVES. 
Three of the seven were observed within 15 minutes of the trigger
by the {\it Swift} satellite. This search has provided high-resolution 
(6.0~\kms\ FWHM) profiles of absorption in the ions \os, \nf, \cf,
\sif, \sfo, and \ssix. The spectra contain a wealth of useful
information on the properties of circumburst and interstellar gas in
the GRB host galaxies. We have presented the results of Voigt
profile fits to the high-ion absorption. Analysis of the line
profiles shows that several types of high-ion absorption exist at
$z\!\approx\!z_{\rm GRB}$, which we summarize in the following points.

\begin{enumerate}
\item 
  We detect strong high-ion absorption components exactly at \zgrb,
  always in \os, \cf, and \sif, usually in \nf, and occasionally
  (where the data are unblended) in \sfo\ and \ssix. The high-ion
  column densities are significantly higher than those seen in the
  Milky Way ISM and in DLAs. We focus on \nf\ since it is the least
  saturated and hence best-measured high ion. In three of seven
  afterglow spectra, the strong \nf\ absorption takes the form of a
  single component coincident in velocity with saturated absorption in
  the other high ions, appearing to support the recent conclusion of
  \citet{Pr08b} that GRBs can photoionize \nf\ in circumburst gas.
  However, in the remaining four cases, the \nf\ absorption is either
  multi-component, offset from the other high-ions, or absent, and it
  is unclear whether the circumburst hypothesis can explain these
  cases. In addition, the \nf\ $b$-values (median of 16~\kms) are not
  narrow enough to rule out collisional ionization. Finally, for 
  GRB~050730 we have placed a lower limit on the distance to the
  \sfo-absorbing gas of 400~pc, based on comparing photo-excitation
  models to the observed column density of \sfo\ and upper limit to
  \sfo$^*$. This supports an interstellar (rather than circumburst)
  origin for the strong high-ion absorption components at \zgrb.

\item In addition to the strong component at \zgrb, the \cf\ and \sif\
  absorption exhibits complex, multi-component profiles extending over
  several hundred \kms\ (excluding the HV components, which we treat
  separately). The median total \cf\ line width in our seven spectra
  is 280~\kms, similar to the value it takes in high-redshift DLA 
  galaxies. This supports a galactic ISM origin for these multiple components.

\item We detect asymmetric, blueshifted absorption-line wings in the 
  \cf, \sif, and (at lower S/N) \os\ profiles in four of the seven GRB 
  afterglow spectra. These wings are similar in shape and velocity
  extent ($\approx$100--150~\kms) to positive-velocity wings seen in
  \os\ absorption at zero redshift in the Milky Way halo, and two of
  the four wings show \cf/\sif\ ratios of $\approx$3, 
  equivalent to the value measured in the Milky Way ISM.
  We thus interpret the wings as tracing outflowing interstellar
  gas. The \os\ column densities in the GRB wings are, in the mean, five
  times higher than the \os\ column densities in the Milky Way wings
  (implying similarly higher mass flow rates), 
  and the gradient d$N_{\rm a}$/d$v$ is typically 5--10 times higher. 

\item HV (500--5000~\kms) components are seen in \cf\ in six of the seven
  spectra (all GRBs except 050922C).
  These components are not necessarily a single homogeneous
  population, and many may arise in unrelated foreground galaxies.
  However, in the (previously unpublished) cases of GRBs~071031 and
  080310, the ionization properties of the HV components (very high
  \cf/\sif\ ratios and an absence of neutral-phase absorption in \siw\
  or \cw) are consistent with a circumburst origin. Models of
  Wolf-Rayet winds from the GRB progenitor stars are capable of
  explaining the kinematics and ionization level of these HV components.
\end{enumerate}

\emph{Acknowledgments} 
We thank the ESO staff for their implementation of the rapid-response mode.
PMV acknowledges the support of the EU under a Marie Curie
Intra-European Fellowship, contract MEIF-CT-2006-041363.
We thank Nicolas Lehner, Chris Howk, and Hsiao-Wen Chen
for comments on the manuscript.

\end{document}